\documentclass[preprint2]{aastex}
\usepackage{graphicx}
\usepackage{xcolor}

\def\msun{M$_{\odot}$}

\slugcomment{Not to appear in Nonlearned J., 45.}

\shorttitle{Luminosity gap and radio properties of BGGs}
\shortauthors{H. Miraghaei }

\begin{document}


\title{No dependence of radio properties of brightest group galaxies on the luminosity gap}

\author{H. Miraghaei\altaffilmark{1},
P. N. Best\altaffilmark{2}, 
R. K. Cochrane\altaffilmark{2,3},
J. Sabater\altaffilmark{2}}

\email{h.miraghaei@maragheh.ac.ir}

\altaffiltext{1}{Research Institute for Astronomy and Astrophysics of Maragha (RIAAM), 
University of Maragheh, Maragheh, Iran}
\altaffiltext{2}{Institute for Astronomy (IfA), University of Edinburgh, Royal 
Observatory, Blackford Hill, EH9 3HJ Edinburgh, U.K.}
\altaffiltext{3}{Harvard-Smithsonian Center for Astrophysics, 60 Garden St. Cambridge, MA 02138, USA}

\begin{abstract}
\noindent

We study the radio and optical properties of the brightest group galaxies (BGGs) in a sample of galaxy groups from the SDSS DR7. The luminosity difference between the BGG and the second ranked galaxy in the group (known as the luminosity, or magnitude, gap) has been used as a probe for the level of galaxy interaction for the BGG within the group. We study the properties of BGGs with magnitude gaps in the range 0-2.7 magnitudes, in order to investigate any relation between luminosity gap and the radio properties of the BGG. In order to eliminate selection biases, we ensure that all variations in stellar mass are accounted for. We then confirm that, at fixed stellar mass, there are no significant variations in the optical properties of the BGGs over the full range of luminosity gaps studied. We compare these optical  results with the EAGLE hydrodynamical simulations and find broad consistency with the observational data. Using EAGLE we also confirm  that no trends begin to arise in the simulated data at luminosity gaps beyond our observational limits. Finally, we find that, at fixed stellar mass, the fraction of BGGs that are radio-loud also shows no trends as a function of luminosity gap. We examine how the BGG offset from the center of group may affect the radio results and find no significant trend for the fraction of radio-loud BGGs with magnitude gap in either the BGG samples with greater or less than 100kpc offset from the center of group.

\end{abstract}


\section{Introduction}
\label{sec:Intro}
\noindent
The radio emission of galaxies at low frequencies is dominated by synchrotron 
emission of free electrons in the interstellar or intergalactic magnetic field. 
Star formation and active galactic nucleus (AGN) activity 
are known to be the primary sources of these 
relativistic electrons. 
The observed radio luminosity of a galaxy is a combination of 
these two processes. Weak micro-Jansky 
radio sources generally have star-forming host galaxies, 
whereas strong radio sources are more commonly hosted by AGN 
(Padovani et~al.\ 2007, 2011; Novak et~al.\ 2018). 
The most powerful radio sources, harbouring supermassive
black holes, may have strong radio jets
which can extend up to Mpcs from the centre of the galaxy.

The simple picture described above is well-established, and a wide 
range of observations have established connections between
radio activity and various galaxy properties.  
The radio emission appears to be tightly correlated with 
the far-infrared (FIR) emission for star-forming galaxies 
(Van der Kruit 1971; G{\"u}rkan et~al.\ 2018), whereas radio-loud AGN 
lie offset from this relation, with more substantial radio emission 
(e.g.\ Hardcastle et~al.\ 2016).
 
In the local Universe, the radio AGN population is dominated by
red, elliptical galaxies with low accretion rates 
(Best \& Heckman 2012; Best et~al.\ 2014). This is 
consistent with the theory of hot accretion flows which
 predict the ubiquity of outflows
in hot accretion mode systems (Narayan \& Yi 1995; Yuan \& Narayan 2014).
The fraction of galaxies that host radio AGN
is a strong function of galaxy stellar mass (Best et~al.\ 2005; 
Brown et~al.\ 2011; Janssen et~al.\ 2012;  
Rees et~al.\ 2016),
black hole mass (Ishibashi et~al.\ 2014, {Bari{\v s}i{\'c}} et~al.\ 2017) and
 optical luminosity (Jiang et~al.\ 2007).
It is also reported that the brightest galaxies within groups or clusters are more
likely to host radio AGN than other galaxies with the same stellar mass 
(Von der Linden et~al.\ 2007; Best et~al.\ 2007). The latter is thought to be 
associated with the enhanced gas cooling within the hot gas halos of groups 
and clusters (Best et~al.\ 2007), but may also reflect an increased importance 
of galaxy interactions in triggering radio AGN (e.g. Sabater et~al.\ 2013). 

The radio luminosity of an AGN is expected to depend fundamentally on
the availability of gas and the efficiency of the accretion of this
gas onto the black hole. The large-scale drivers of these dependencies can 
be studied observationally via investigating the properties and environment of the host galaxy.
However, there may also be some intrinsic 
properties influencing the powering of radio AGN, e.g. the black hole spin
 (Garofalo et~al.\ 2010, McNamara et~al.\ 2011),  
that are not detectable using current 
observational facilities.

The relation between radio properties and galaxy environment, 
in particular in terms of interactions and mergers, has been studied extensively.
These include investigations of the radio power in
major and minor merging systems (Chiaberge et~al.\ 2015),
 non-merging systems (Miraghaei et~al.\ 2014),
 dense environments (Sabater et~al.\ 2013, Kolwa et~al.\ 2018) 
or galaxy clusters (Mittal et~al.\ 2009, Mo et~al.\ 2018),
 galaxy pairs ({Argudo-Fern{\'a}ndez} et~al.\ 2016)
and isolated galaxies (Sabater et~al.\ 2008). 
These studies have provided mixed results, with some in 
favour of a positive effect of galaxy interactions
on triggering AGN while others report no enhancement in AGN activity due to galaxy interactions 
(Coziol et~al.\ 2017, Wen et~al.\ 2012). The relations derived are sensitive to the 
type of AGN studied, i.e. optical, IR or radio AGN 
(Ellison et~al.\ 2008, 2011, 2015; Satyapal et~al.\ 2014; Miraghaei 2020) or the rate of gas accretion into the 
super massive black hole as discussed earlier. The latter is probed using a sample of high 
excitation or low excitation AGN (Sabater et~al.\ 2012; 2013; 2015). 
The morphology of galaxies studied can also play a role (Davies et~al.\ 2017). 
Furthermore, the very strong dependence of the radio-AGN fraction on stellar 
mass means that sample selection biases can have a profound effect on the results obtained. 

In this paper, we study the effect of galaxy interactions on radio-mode AGN activity 
by exploring the properties of the brightest group galaxy (BGG) 
in galaxy groups as a function of the difference between 
the optical luminosities of the first and second 
ranked galaxies within a group: the luminosity gap. 
The BGG in a galaxy group is formed by the continuous merging of smaller galaxies, 
which can lead to a large gap between the luminosities of the first and second 
ranked galaxies in the group. The brighter galaxies in a group have a relatively shorter merging time-scale
and so a galaxy system with a small luminosity gap is more likely to undergo a merger. 
On the contrary, galaxy groups with large luminosity gaps,
 which tend to comprise one dominant galaxy and some dwarf galaxies with 
long merging timescales, undergo fewer interactions and are more likely to be merger-free. 
(Lacey \& Cole 1993; van den Bosch et~al.\ 2007; Jiang et~al.\ 2008). A study by Dariush
et~al.\ (2007) also shows that galaxy groups with magnitude gaps greater 
than 2 display a lack of major mergers over the past 4 Gyr of their histories 
(compared to groups without a large magnitude gap). Using the Millennium simulation, they show that these systems 
are early-forming, having accumulated 80$\%$ of their mass before this time.
This may have observational consequences on the 
properties of the central galaxies in the groups, which is the subject of our work. 

The radio luminosities of large magnitude gap systems ($\Delta$M$\rm_{r}$$\geq$1.7) have been studied 
by Khosroshahi et~al.\ (2017). They showed that BGGs that reside within 100kpc of 
the centre of a galaxy group with a large magnitude gap are under-luminous (in the 
radio) compared to BGGs that lie more than 100kpc from the centre of small magnitude gap systems. 
To investigate this further, we explore how the magnitude gap correlates with the 
radio-mode AGN activity in galaxy groups, focussing on the BGG. 
Our aim is then to understand the opposing trends 
observed for small and large magnitude gap systems, using a large, continuous range in r-band
magnitude gap, from $\Delta$M$\rm_{r}$=0 to $\Delta$M$\rm_{r}$=2.7.
In addition, we investigate whether an offset in the position of the BGG from the center of 
the group would affect our results. 
We are particularly careful to avoid 
potential selection biases due to stellar mass effects. To achieve this, 
we draw a large sample of galaxy groups from the
seventh data release of the Sloan Digital Sky Server (SDSS), and then make 
conservative selection cuts to ensure unbiased samples. 
The sample has been limited to radio-loud AGN and 
corrected for the radio emission from star formation.

To confirm that our results are not influenced by any 
secondary effects, we also investigate any dependence of the optical properties of the BGGs and 
the magnitude gap. In this regard, Trevisian et al. (2017) find no significant correlation 
between the optical properties and the magnitude gap up to $\Delta$M$\rm_{r}$=2.5.
We further compare the optical properties of the BGGs as a function of 
luminosity gap to predictions 
from state-of-the-art cosmological simulations, both to support the 
observational results and to ensure that the simulations do not predict 
any change in behaviour at luminosity gaps larger than we are able to 
probe with our unbiased observational samples.
We use the Evolution and Assembly 
of GaLaxies and their Environments (EAGLE) simulation 
(Schaye et al.\ 2015, Crain et al.\ 2015), which follows the formation 
of galaxies and supermassive black holes in a standard $\Lambda$CDM universe.  
The simulation includes subgrid prescriptions for radiative cooling, 
star formation, stellar evolution, mass-loss and metal
enrichment, energy feedback from star formation and AGN, mergers, gas accretion 
and the growth of supermassive black holes. Feedback from
star formation and AGN is implemented thermally, such that outflows
 develop as a result of pressure gradients and without the need
to impose winds by hand. The simulation reproduces observational properties 
of galaxies with an accuracy that is unprecedented for hydrodynamical simulations.

The layout of this paper is as follows. The galaxy and radio source samples
 are presented in Section 2. Section 3 presents the construction of matched sub-samples.
The connection between luminosity gap and optical and radio properties of the BGGs are 
shown in Section 4 and 5 respectively. In Section 6, we consider the impact 
of a spatial offset of the BGG from the centre of the group on the 
radio results presented in Section 5.  We summarise and draw
conclusions in Section \ref{sec:summery}.  Throughout the paper we assume
a $\Lambda$CDM cosmology with the following parameters: $\Omega_m=0.3$,
$\Omega_\Lambda=0.7$ and $H_0=100 h$ km s$^{-1}$ Mpc$^{-1}$ where $h$ =
0.70.\\

\section {Sample and classification }
\label{sec:Sample}

\begin{figure*}
\centering
 \includegraphics[ scale=0.45]{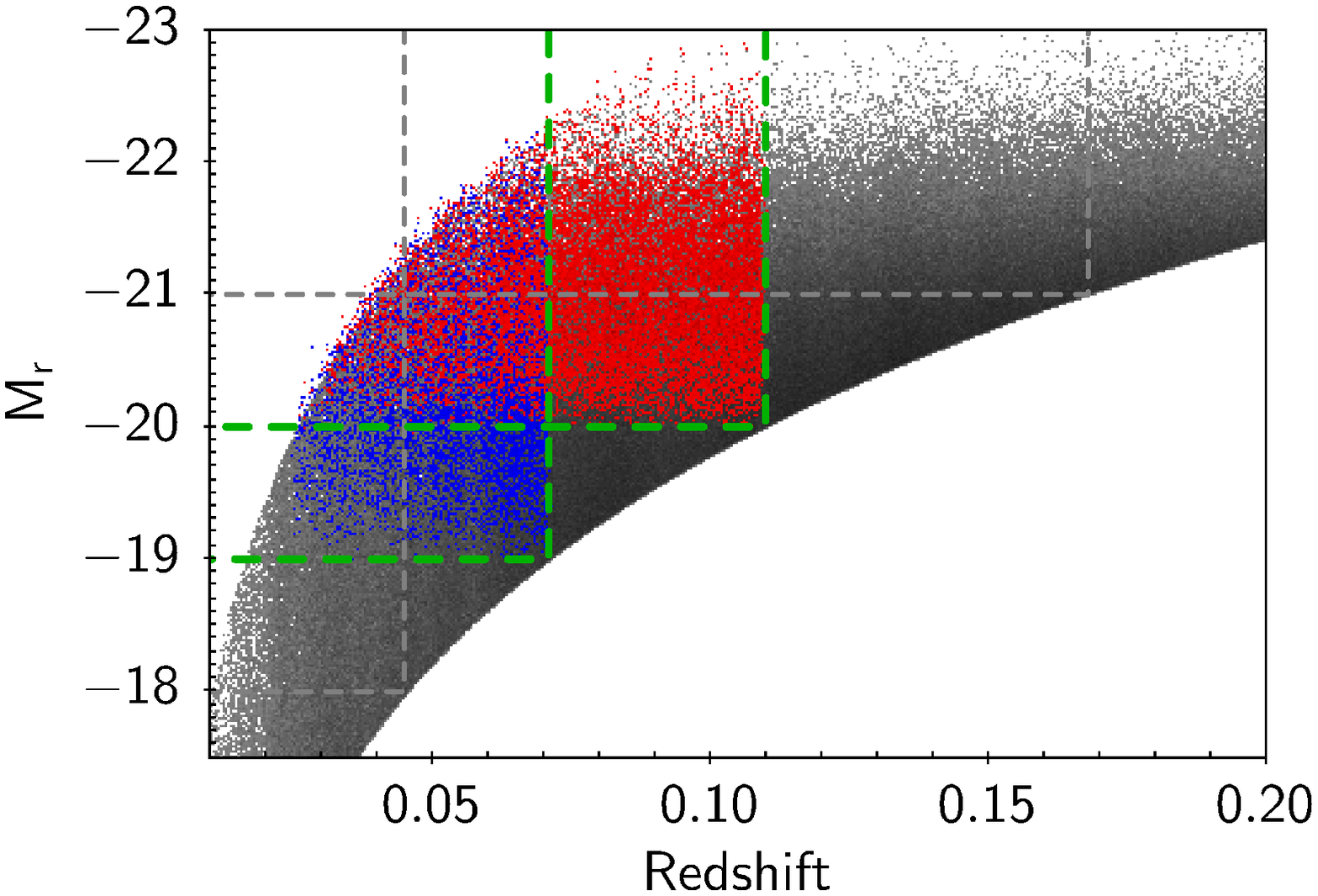}
 \includegraphics[ scale=0.45]{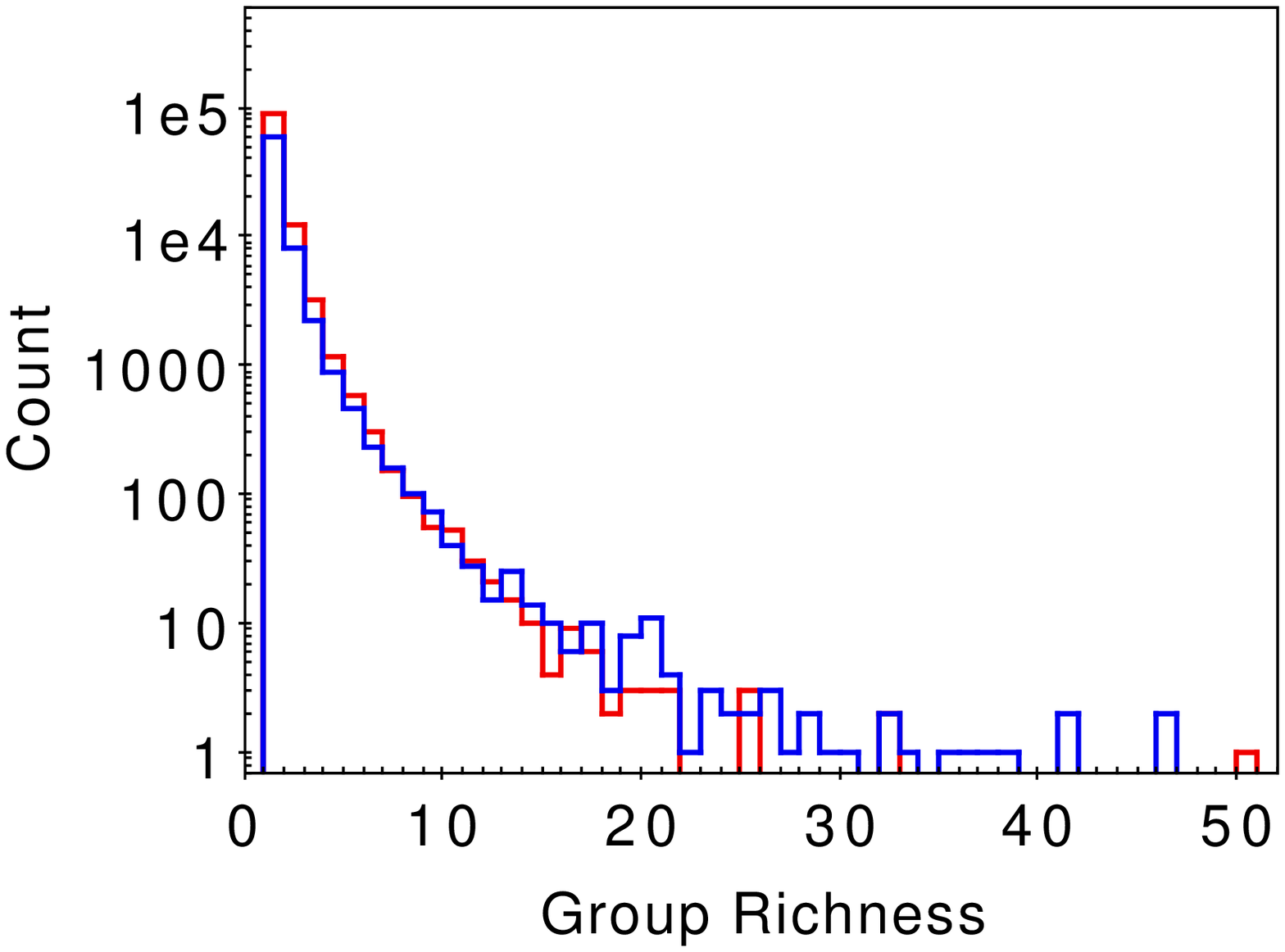}
\figcaption{The left panel shows r-band absolute magnitude versus redshift for 
the SDSS main galaxy sample (grey) along with the Tago et~al.\ (2010) BGGs with 
group members brighter than r-band absolute magnitude cuts of -19 (blue) and -20 (red).
The green dashed lines show the limits in redshifts and magnitudes of the BGGs considered in this study.
The grey dashed lines show the samples which were available in Tago et~al.\ (2010) and excluded
in this study.
The right panel shows number of galaxy members (richness) in the Tago et~al.\ group samples
including the number of isolated galaxies (richness=1) for M$\rm_{r}$=$-$19 (blue) 
and M$\rm_{r}$=$-$20 (red) cuts presented in the left panel.
\label{figsample}}
 \end{figure*}

\noindent
The galaxy sample and the radio and optical properties are taken from Best
\& Heckman (2012). They cross-matched the seventh data release (DR7;
Abazajian et~al.\ 2009) of the Sloan Digital Sky Survey (SDSS; York
et~al.\ 2000) with the National Radio Astronomy Observatory (NRAO) Very
Large Array (VLA) Sky Survey (NVSS; Condon et~al.\ 1998) and the Faint
Images of the Radio Sky at Twenty centimetres (FIRST) survey (Becker,
White \& Helfand 1995), following the techniques of Best et~al. (2005a).
We use the SDSS Main Galaxy Sample, which is a 
magnitude-limited sample with 14.5$<$m$_{r}$$<$17.77. 

Parameters that we use from the catalogue include total stellar mass, black hole mass
(estimated from the velocity dispersion using the relation of 
McConnell \& Ma 2013), 4000$\AA$ break strength as a probe for the stellar age,
galaxy magnitude, 
rest-frame g-r colour (which we convert to the AB magnitude system),
 half light radius in the i-band (R$_{50}$) as a proxy for galaxy size, concentration: 
C=R$_{90}$/R$_{50}$, half-light surface mass density:
$\mu_{50}$ = 0.5 M$_{\star}$/ ($\pi$
  R$_{50}$$^{2}$), 
 and radio luminosity including the radio AGN/star forming galaxy (SFG) separation 
(see Miraghaei \& Best 2017 for the details). We used the rest-frame 1.4 GHz radio luminosities
calculated from NVSS flux densities, assuming ${\alpha}$=0.75 for the spectral index
(where $S_\nu \propto \nu^{-\alpha}$).

To identify galaxies in groups against field galaxies, 
we used the Tago et~al.\ (2010) group samples which provide 
volume-limited group samples of SDSS galaxies, based on 
the modified friends-of-friends (FOF) algorithm.
The limits of M$_{r}$~=~$\mathcal{M}$$_{r}$$-$5~log(h)~$<$~-18, -19, -20 and -21  
(where $\mathcal{M}$$_{r}$ is the r-band absolute magnitude)
 were used by Tago et al to construct 
four group samples; these correspond to the redshifts of 
0.045, 0.071, 0.110 and 0.168 at the magnitude limit 
of the SDSS main galaxy sample (see Fig. \ref{figsample}, left). 
The galaxy group catalogues provide group member locations in
addition to their redshifts and the SDSS optical magnitudes. 
The group center was defined as the center of light derived from the r-band luminosity 
of the group members (Robotham et~al.\ 2011). The projected comoving distance from 
the center of the group was defined as the {\it offset} from the center for each of the group members. 
The r-band magnitude gaps, $\Delta$M$\rm_{r}$, were calculated for galaxy groups 
with two or more galaxy members within 500 kpc h$^{-1}$ radius. 
The rest of the galaxies were classified as isolated galaxies.

The four different group sample cuts from Tago et~al were considered and the  
-19 and -20 cuts were selected to present in this work.
A -18 magnitude cut did not provide sufficient volume (number statistics) 
while a -21 magnitude cut does not allow a sufficient range of luminosity gap.
 Either of the group sample with -19 and -20 cuts provide 
good sample sizes. The former has better ability to reach higher magnitude gap 
for the lower redshift galaxy groups and the latter gives better statistics at lower magnitude
gap for the higher redshift galaxy groups; hence the results 
from both are considered. The distribution of the group richness
is presented in Fig. \ref{figsample} (right).

\section {Construction of matched sub-samples}
\label{sec:R}

\begin{figure*}
\centering
 \includegraphics[ scale=0.3]{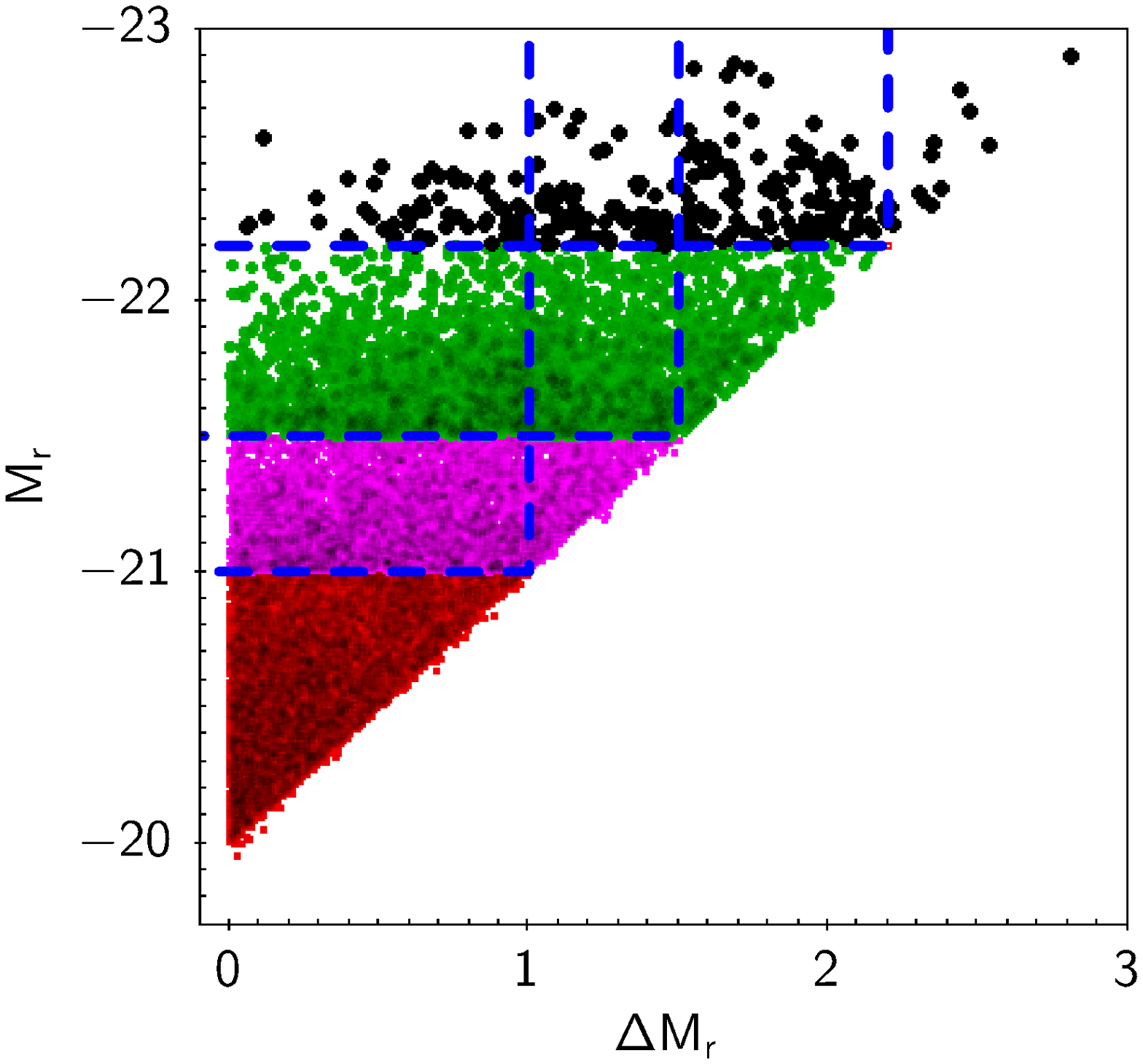}
 \includegraphics[ scale=0.3]{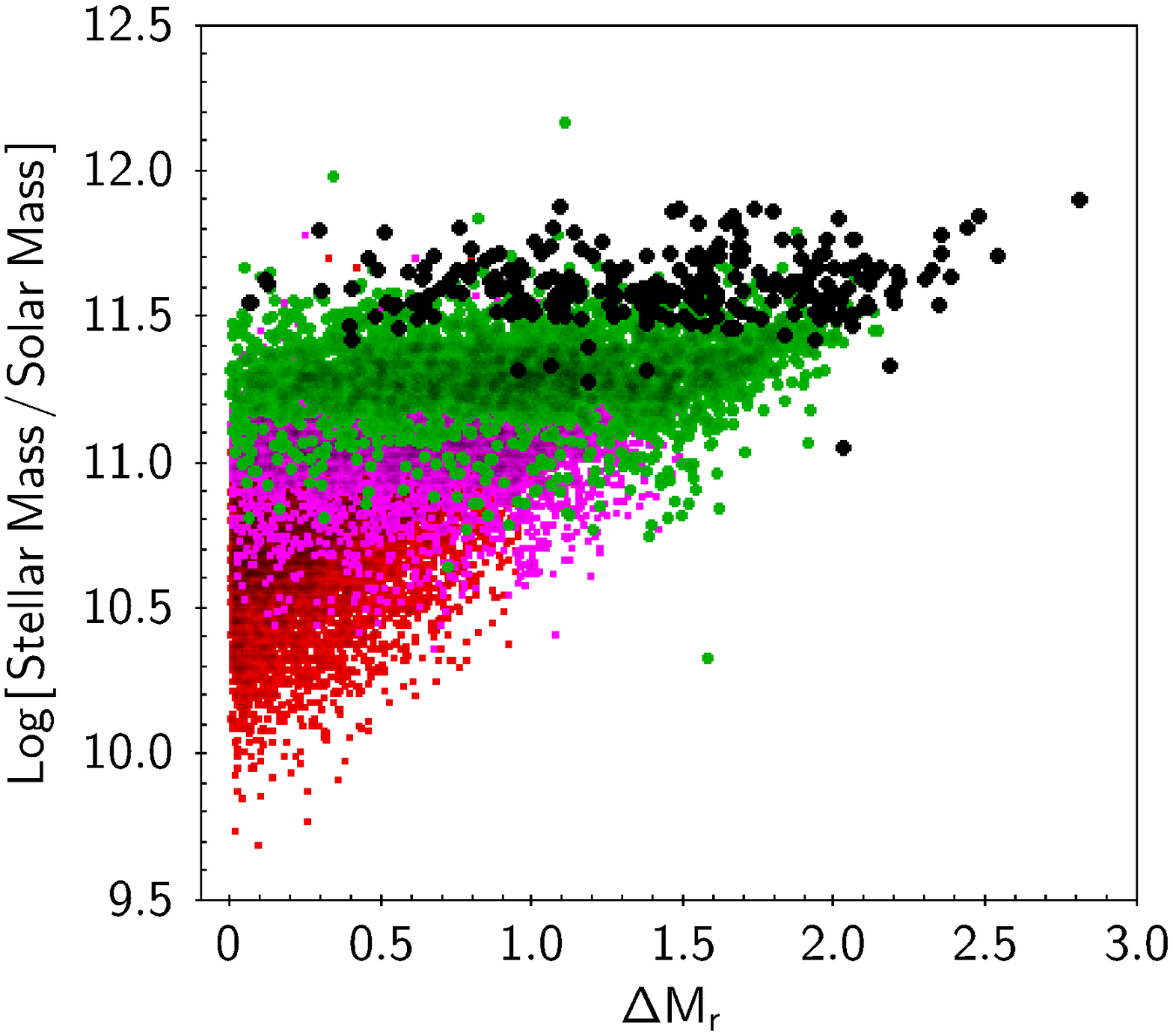}
 \includegraphics[ scale=0.3]{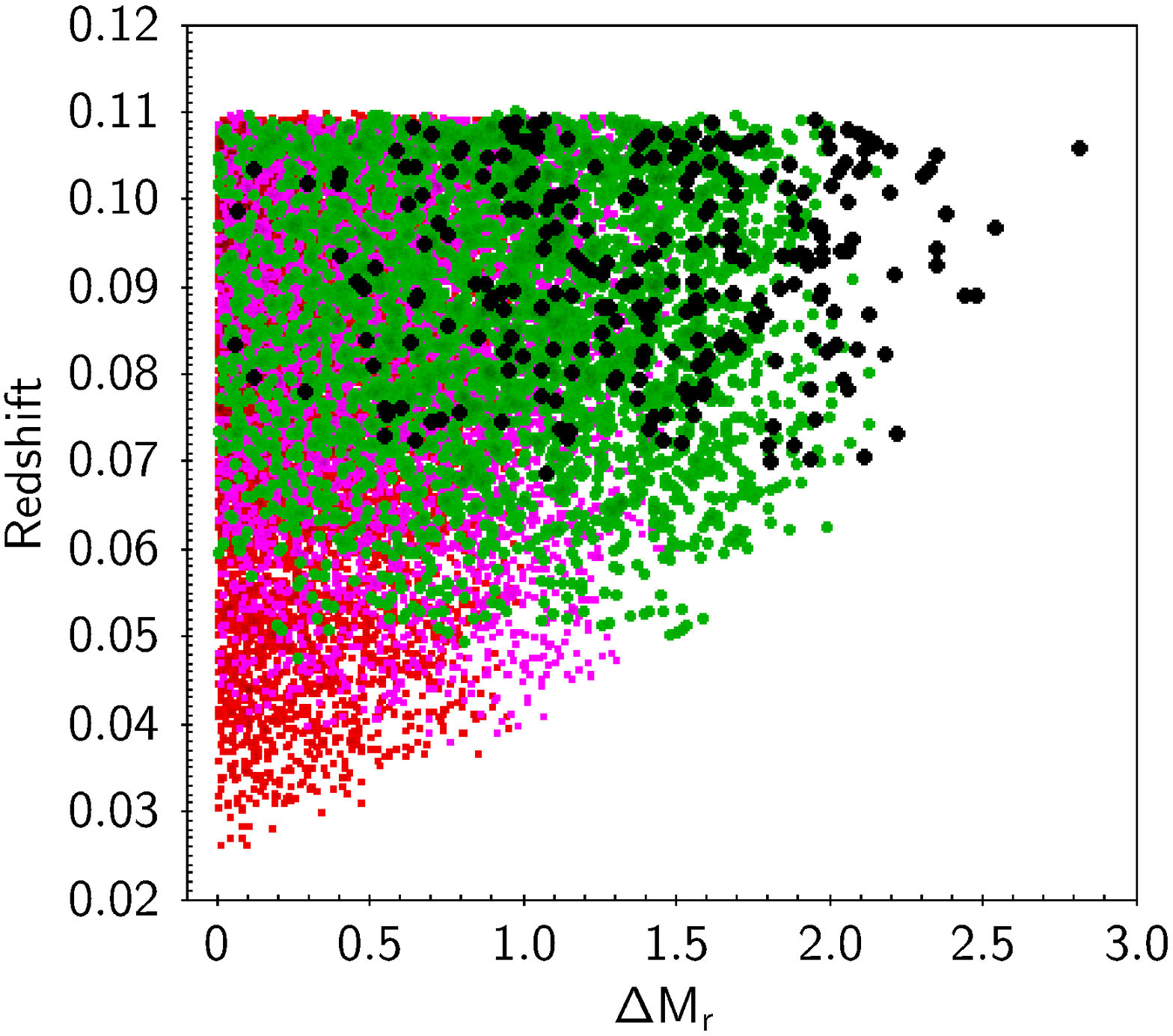}
 \includegraphics[ scale=0.3]{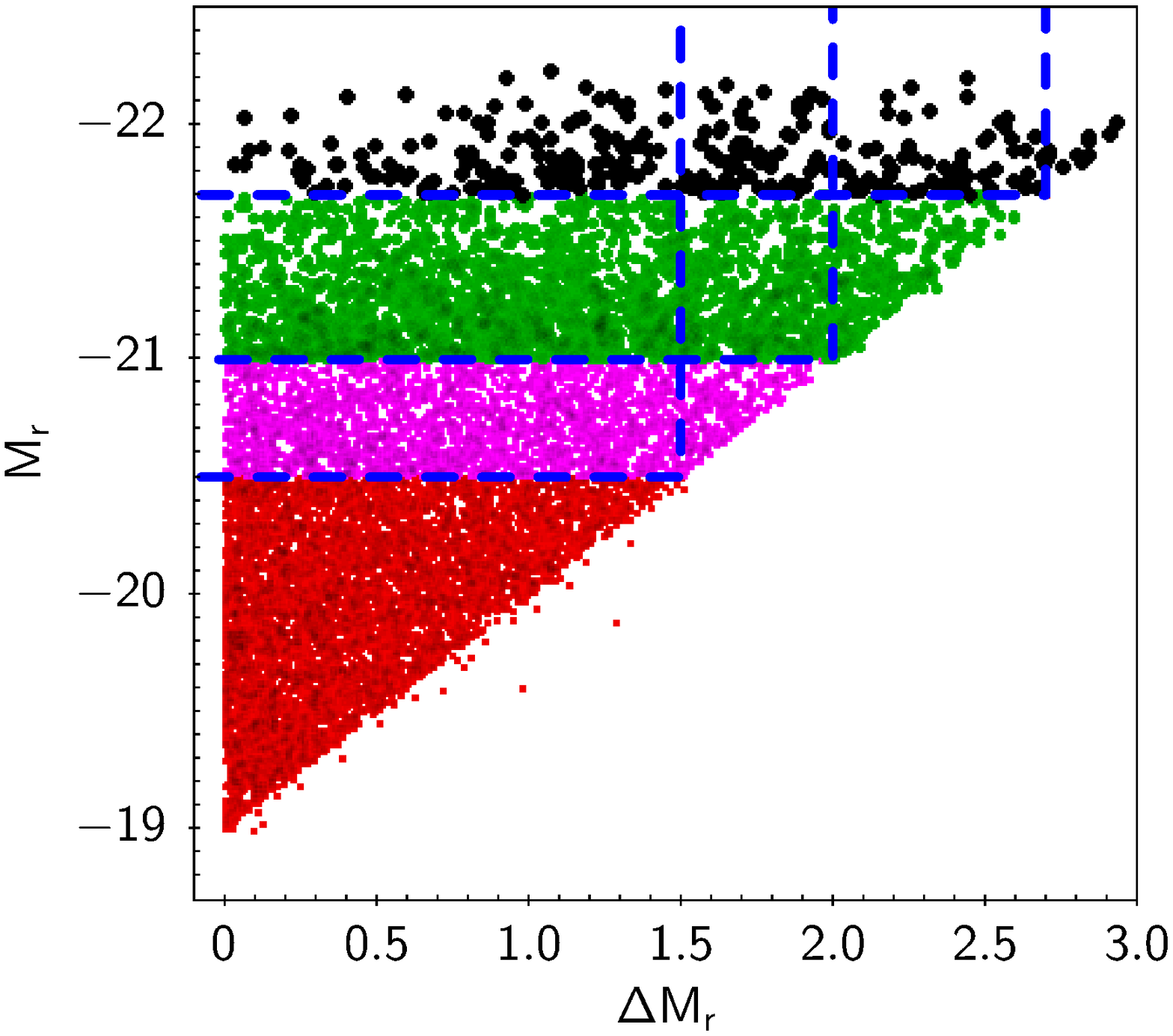}
 \includegraphics[ scale=0.3]{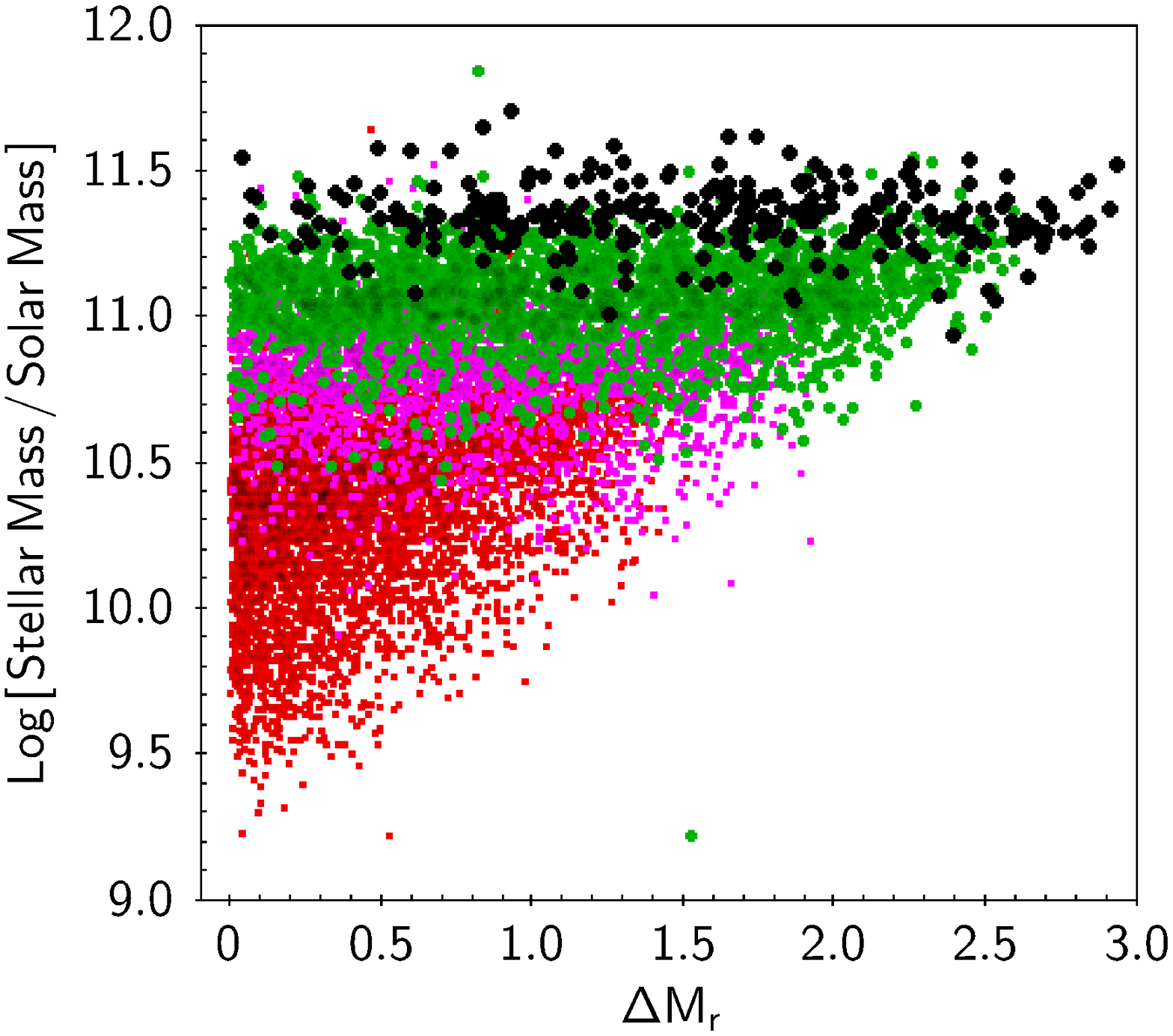}
 \includegraphics[ scale=0.3]{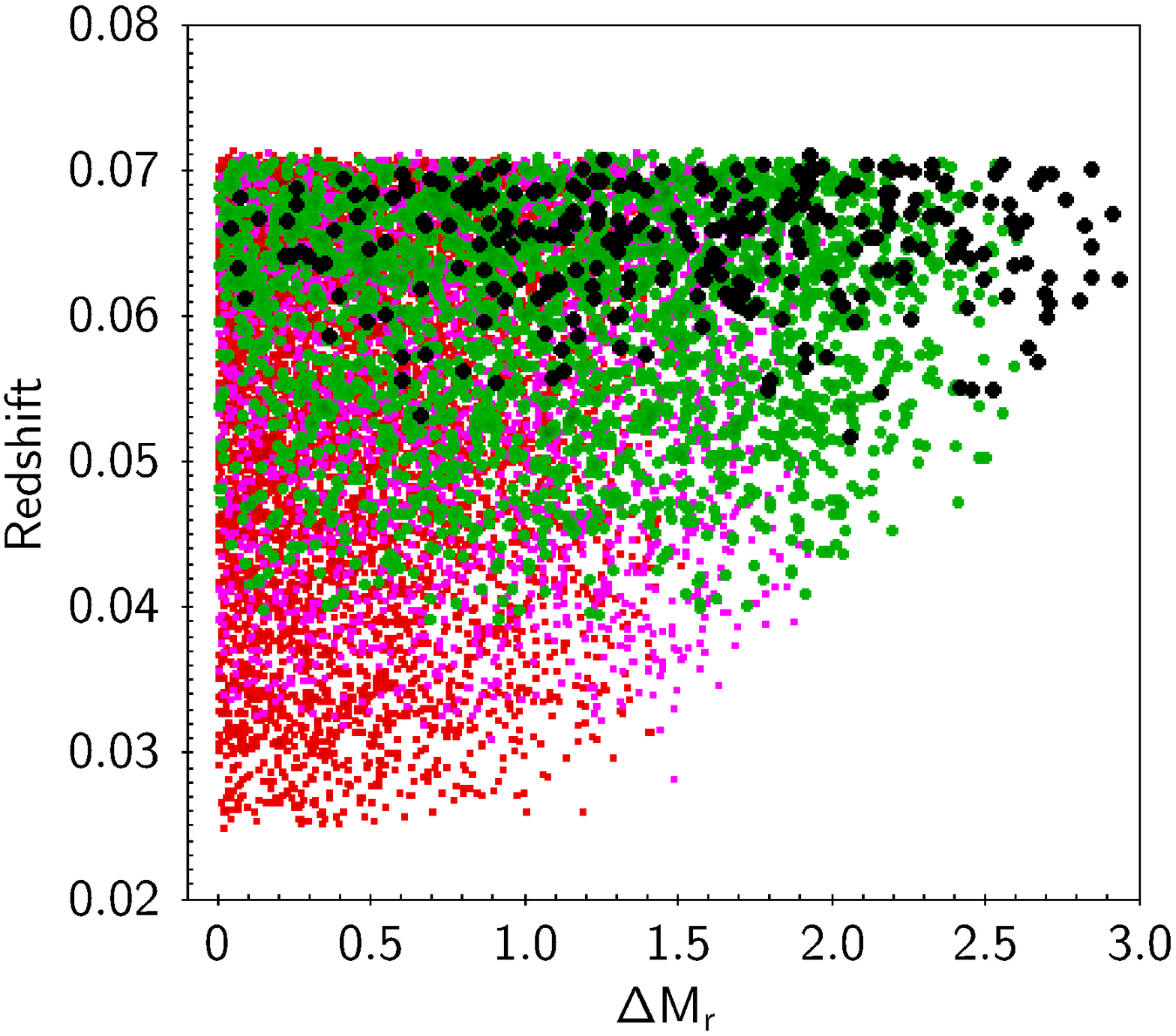}
\figcaption{The r-band absolute magnitude (left), stellar mass (middle) and 
redshift (right) versus magnitude gap for 
the BGG samples with group member absolute magnitude cuts of -20 (upper panels) 
and -19 (lower panels) presented in Fig. \ref{figsample}. The different colours indicate three different sub-samples, with 
-21 (pink), -21.5 (green) and -22.2 (black) cuts in BGG r-band absolute magnitude for the -20 cut sample and -20.5 (pink), 
-21 (green) and -21.7 (black) cuts in BGG r-band absolute magnitude for the -19 cut sample. 
The dashed lines in the left-hand panel show
up to which magnitude gaps the sub-samples are reliable.
\label{figHist1}}
 \end{figure*}

\noindent
To investigate the radio and optical properties of the BGGs, it is 
first essential to ensure that we are making unbiased comparisons of the BGGs 
with different luminosity gaps. The r-band magnitudes, stellar masses and
redshifts of the BGGs for each sample are shown in Fig.~2 (red circles). 
Galaxy groups with large luminosity gaps have BGGs with brighter r-band 
magnitudes, higher stellar masses, and are typically located 
at higher redshifts (the last of which arises from the correlation between 
stellar mass and redshift  due to selection biases in the magnitude-limited 
SDSS sample). This dependence arises trivially due to the r-band  
magnitude -19 (or -20) limit of the Tago et~al.\ (2010) group 
sample. Accordingly, a system can only be identified as a group if the 
second-ranked galaxy within the group has a magnitude of -19 (or -20) or brighter, 
and hence at large luminosity gaps only BGGs of the brightest magnitudes are 
included in the catalogue. BGGs with large luminosity gaps at fainter 
magnitudes will be identified as isolated galaxies which imposes incompleteness 
on the catalogue at large luminosity gaps. This incompleteness means that any naive comparison
of the BGGs at different luminosity gap would be biased by 
differences in stellar mass, magnitude and redshift, 
which could easily have more influence than the luminosity gap itself.

To correct for
this effect, we build up three subsamples with additional magnitude cuts
on the BGG. This is illustrated in the left panel of Fig.~2.  
For the galaxy group sample with -19 cut (lower left), a second 
cut at M$\rm_{r}$=-20.5
for the BGGs allows for a complete sample out to $\Delta$M$\rm_{r}$=1.5, as do cuts  
at  M$\rm_{r}$=-21.0 and  M$\rm_{r}$=-21.7 out to $\Delta$M$\rm_{r}$=2.0 and
$\Delta$M$\rm_{r}$=2.7 respectively. In the -20 cut sample 
(upper left), the second cuts for the BGGs 
at M$\rm_{r}$=-21.0, -21.5 and -22.2 allow for complete samples 
out to  $\Delta$M$\rm_{r}$=1.0, 1.5 and 2.2 respectively. 
These cuts also largely remove the mass and
redshift biases in the distribution of the BGGs at different luminosity 
gap  for each of the samples up to their completeness limits 
for $\Delta$M$\rm_{r}$.  The stellar mass (middle panel) 
and redshift (right panel) distributions are
shown in Fig. 2. Nevertheless, we apply additional mass limits to each subsample, 
derived from the middle panel of Fig.~2 to include the 
bulk of the BGGs in the given magnitude range, while 
ensuring that the results are not dominated by any outlying sources.

Specifically, in the rest of the paper, we study
three sub-samples of the BGGs with three different 
cuts on BGG absolute magnitude and in specific mass ranges 
for each of the -20 and -19 galaxy group samples.
\newline
\newline
\noindent
A: Galaxy group sample with -20 cut:
\begin{description}
\item   i-  M$\rm_{r}$$\leq$-21.0, 10.9$\leq$Log$(M{*}/M_{\odot})$$<$11.2, $\Delta$M$\rm_{r}$$\leq$1.0
\item   ii- M$\rm_{r}$$\leq$-21.5, 11.2$\leq$Log$(M{*}/M_{\odot})$$<$11.5, $\Delta$M$\rm_{r}$$\leq$1.5 
\item   iii-M$\rm_{r}$$\leq$-22.2, 11.5$\leq$Log$(M{*}/M_{\odot})$$<$11.9, $\Delta$M$\rm_{r}$$\leq$2.2 
\end{description}

\noindent
B: Galaxy group sample with -19 cut:
\begin{description}
\item   i-  M$\rm_{r}$$\leq$-20.5, 10.7$\leq$Log$(M{*}/M_{\odot})$$<$11.0, $\Delta$M$\rm_{r}$$\leq$1.5
\item   ii- M$\rm_{r}$$\leq$-21.0, 11.0$\leq$Log$(M{*}/M_{\odot})$$<$11.3, $\Delta$M$\rm_{r}$$\leq$2.0 
\item   iii-M$\rm_{r}$$\leq$-21.7, 11.3$\leq$Log$(M{*}/M_{\odot})$$<$11.7, $\Delta$M$\rm_{r}$$\leq$2.7 
\end{description}

These selection criteria allow the investigation of the largest
complete and unbiased samples of BGGs up to the specific magnitude
gap. In addition, we reduce the uncertainties which arise from 
galaxy group detection by making a cut on the group 
richness. The FOF group finding algorithm links every galaxy to those
neighbouring galaxies and their neighbours by defining a length scale as a maximum distance between the two
galaxies (Huchra \& Geller 1982). Although this method is widely used for 
identification of galaxy groups in local redshift surveys, the algorithm can not identify the
physical state of the group. Therefore, spurious groups generated 
from the chance alignment of physically unrelated galaxies, 
or infalling galaxies in a structure which is not yet virialised, may be selected as
true galaxy groups. We discuss this further in Section \ref{sec:off}. We consider galaxy group samples with 
richness of N $\geq$ 2, 3, 4 and 5 throughout the paper. The N $\geq$ 2 selection allows 
the largest samples, and minimises biases with respect to magnitude gap, but a galaxy group with 
only two galaxy members that satisfy the magnitude cut
 used in this work may be a simple galaxy pair, rather than a galaxy group. 
In contrast, the higher N cuts allow us to study richer groups 
 but at the cost of 
smaller sample statistics and potential small biases at larger 
magnitude gaps (since the Nth-ranked group member has to be 
brighter than the SDSS magnitude limit). 
 
\begin{figure*}
\centering
 \includegraphics[ scale=0.3]{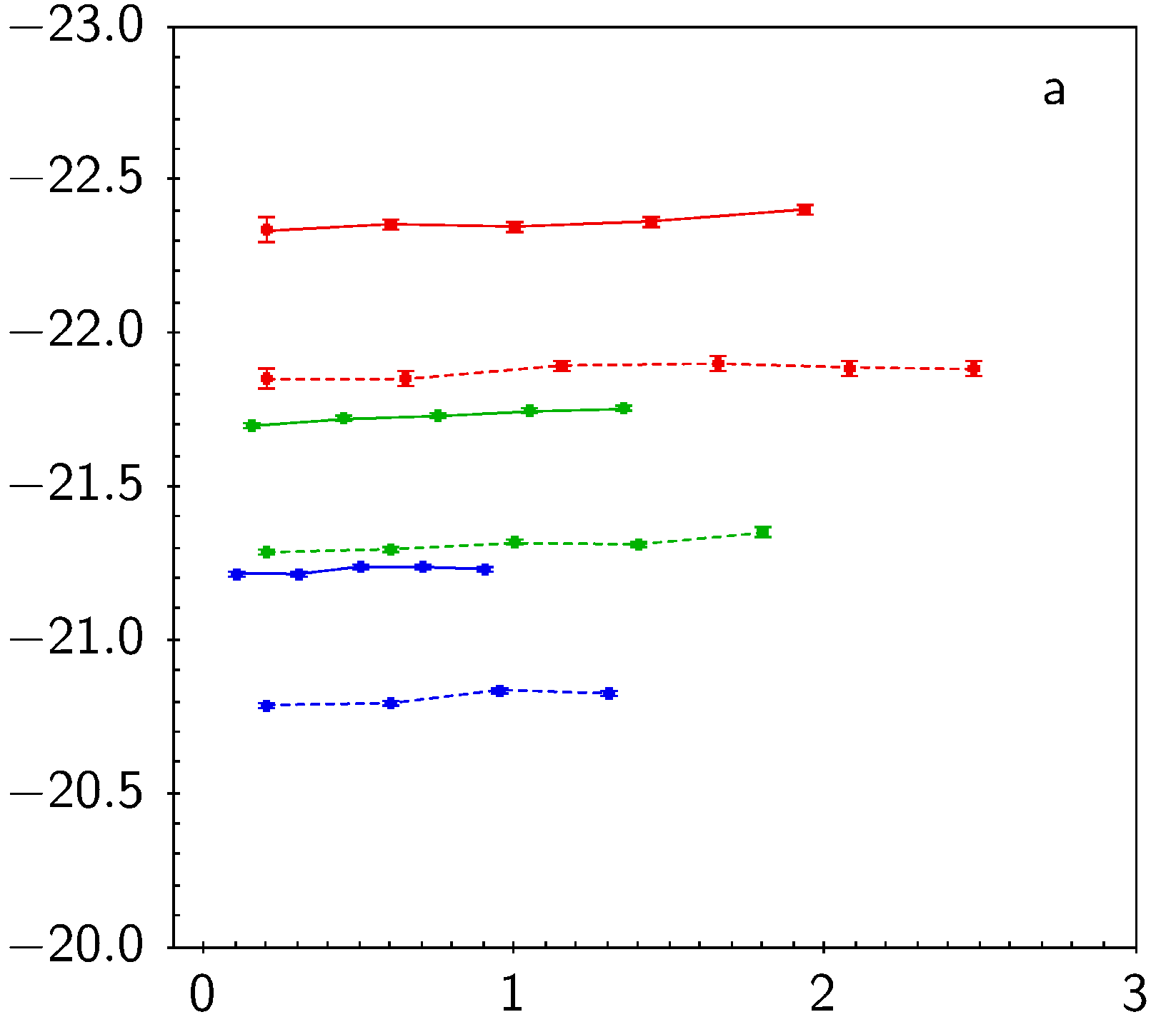}
 \includegraphics[ scale=0.3]{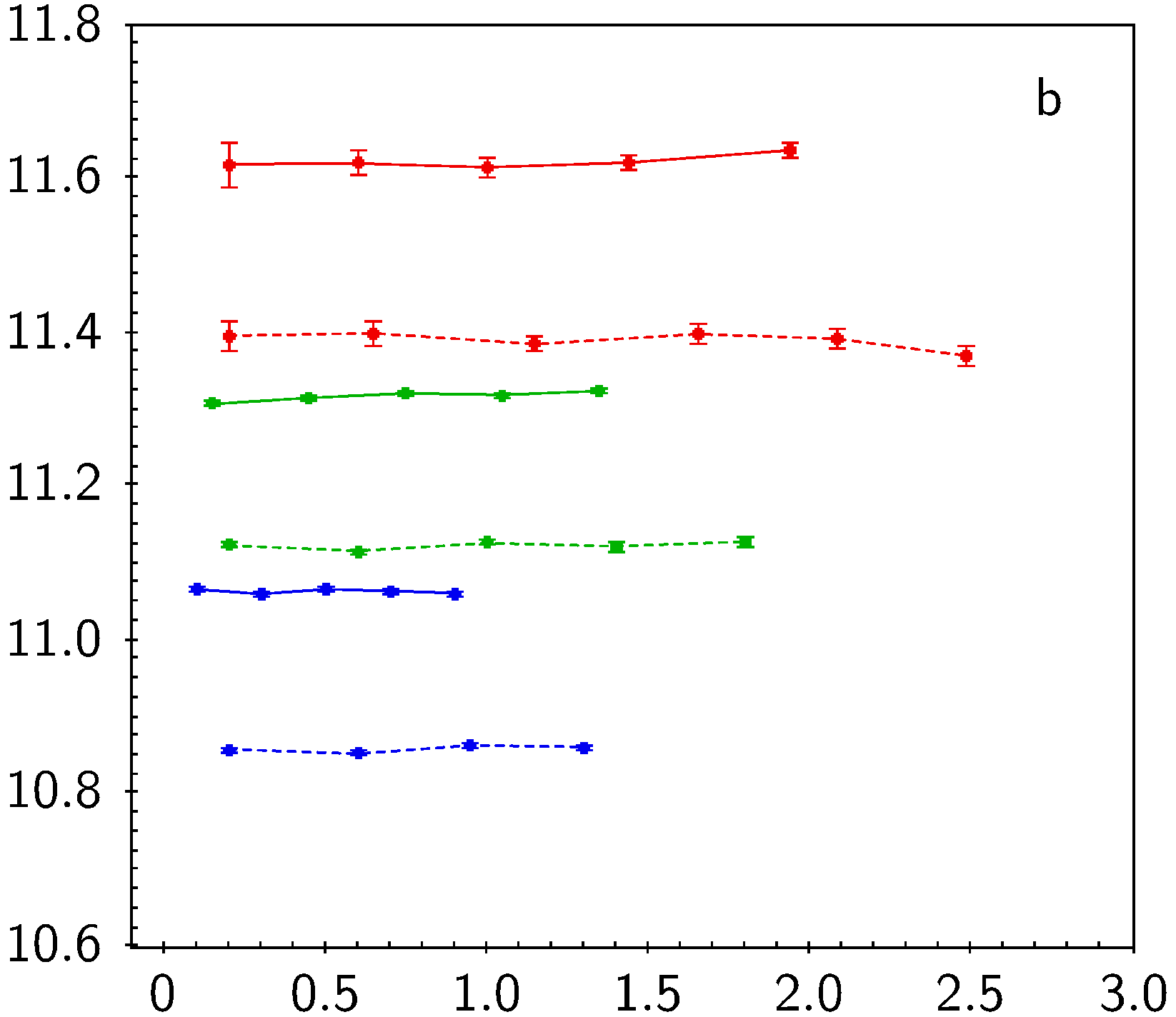}
 \includegraphics[ scale=0.3]{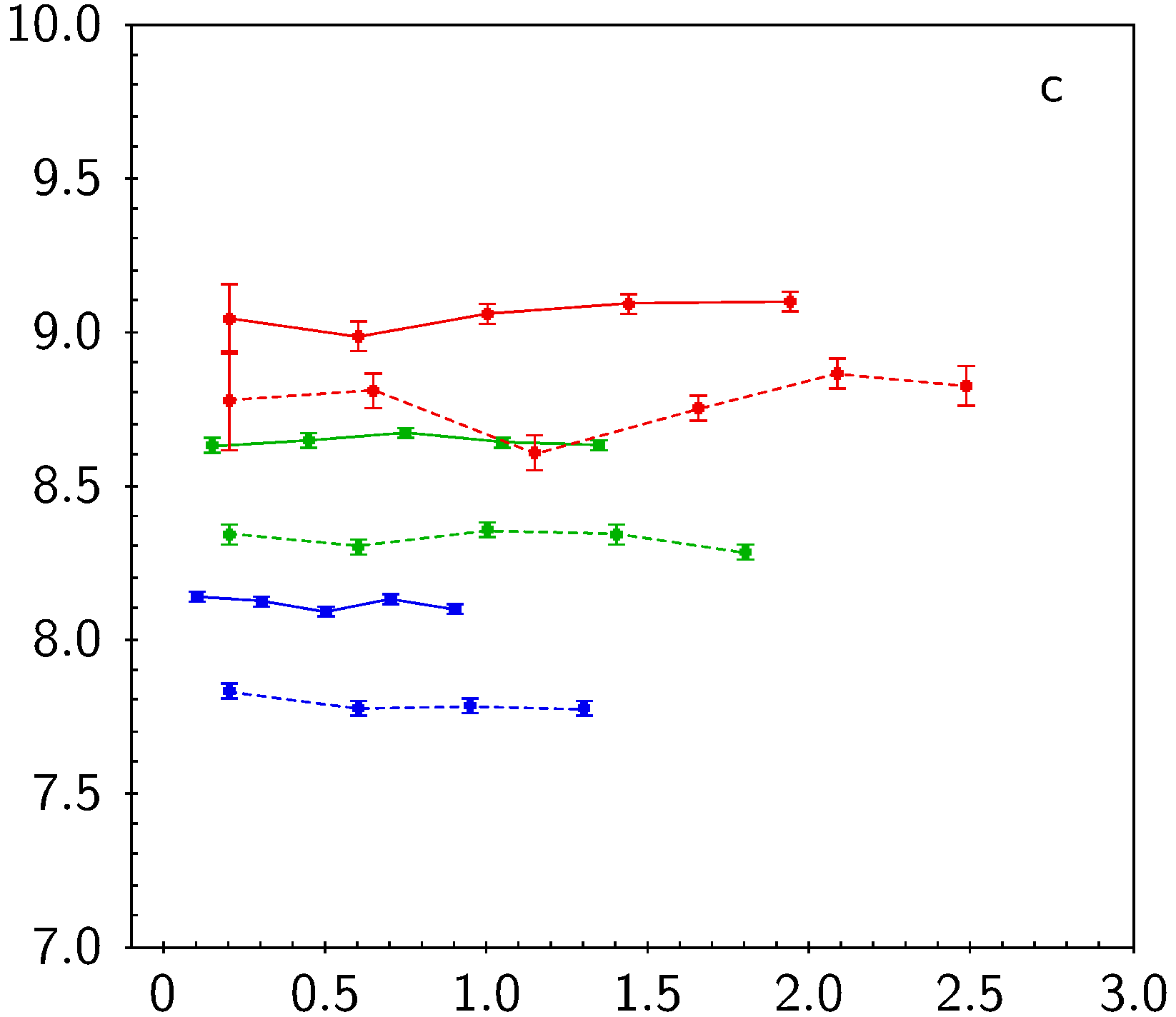}
 \includegraphics[ scale=0.3]{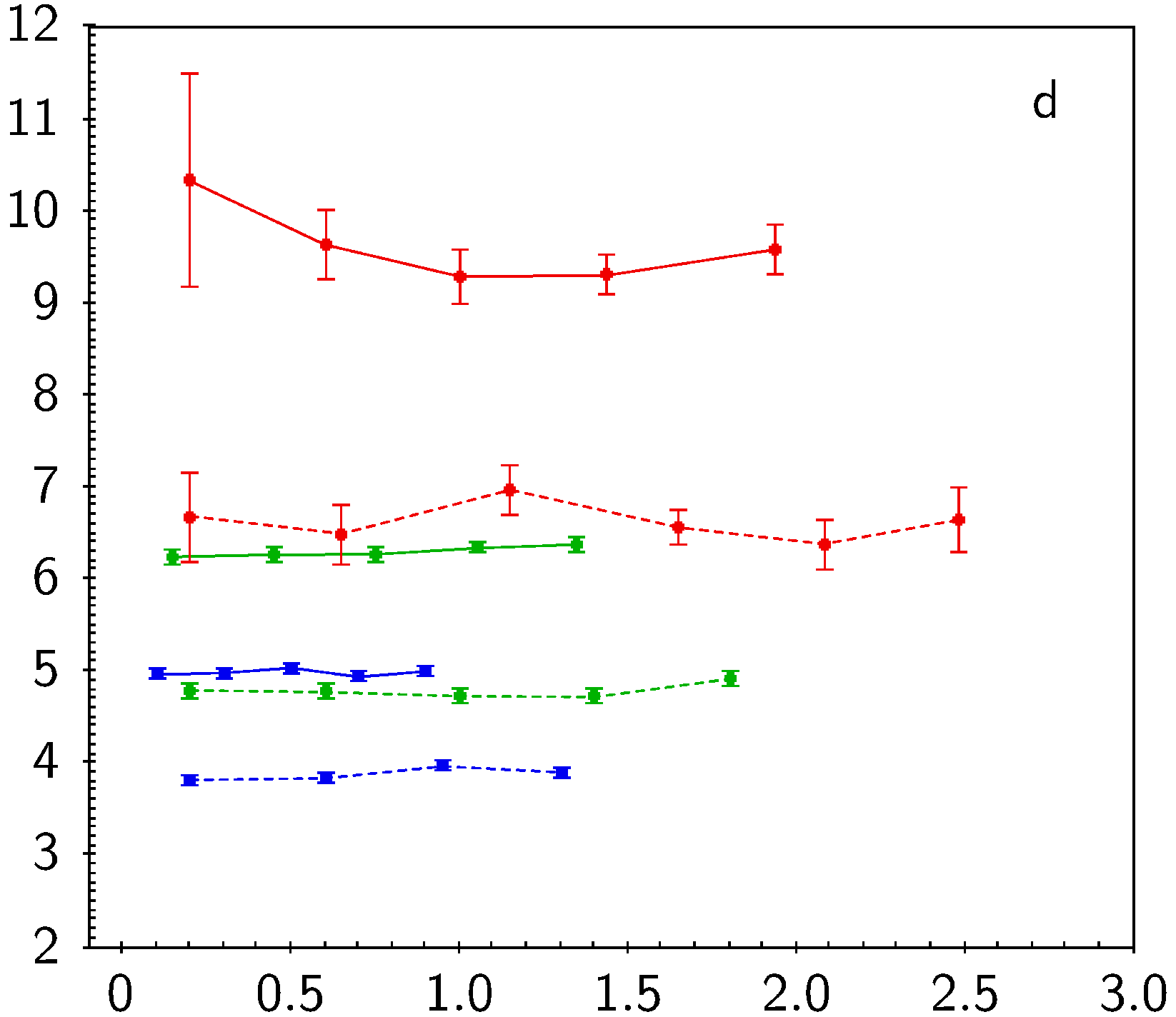}
 \includegraphics[ scale=0.3]{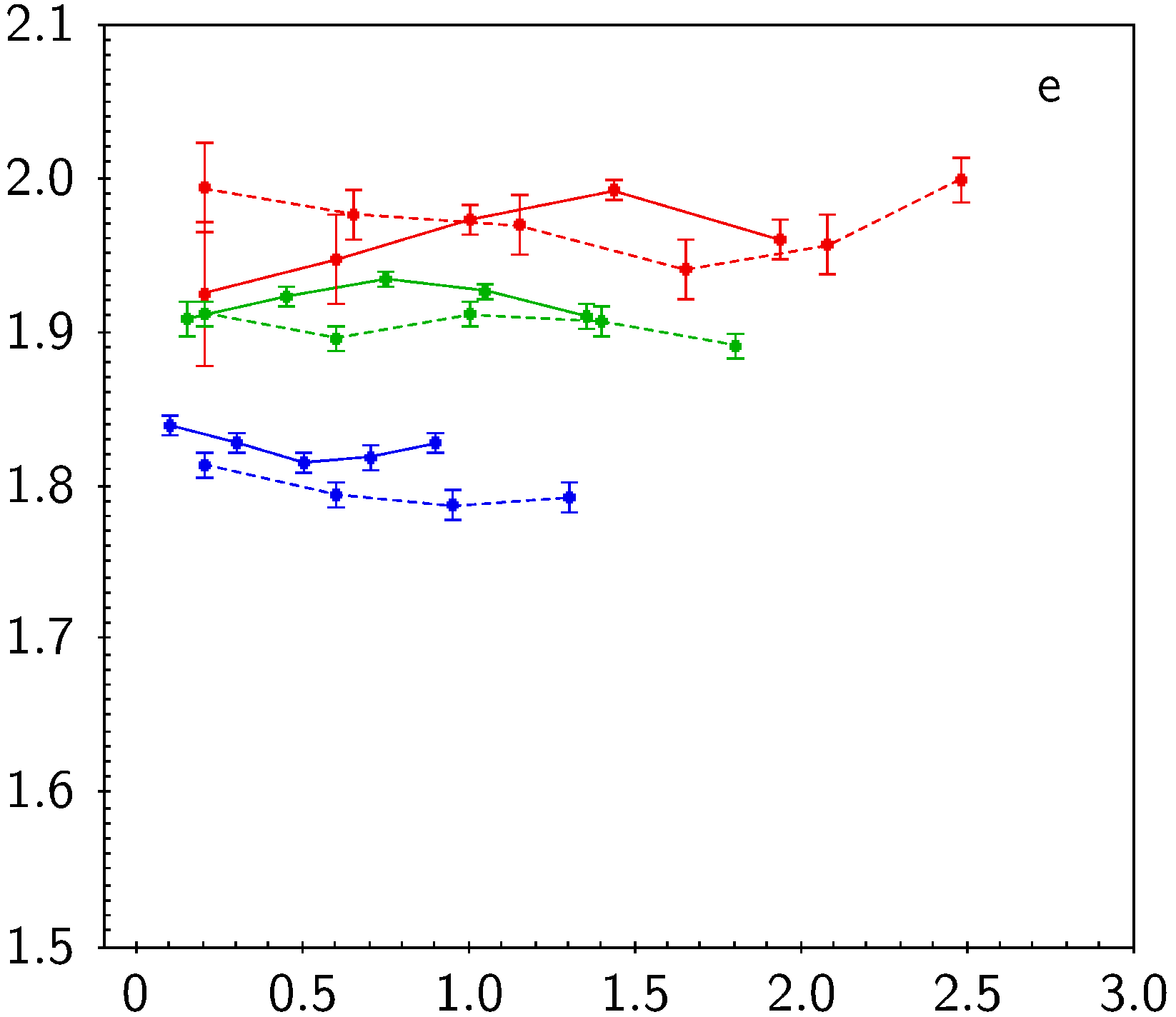}
 \includegraphics[ scale=0.3]{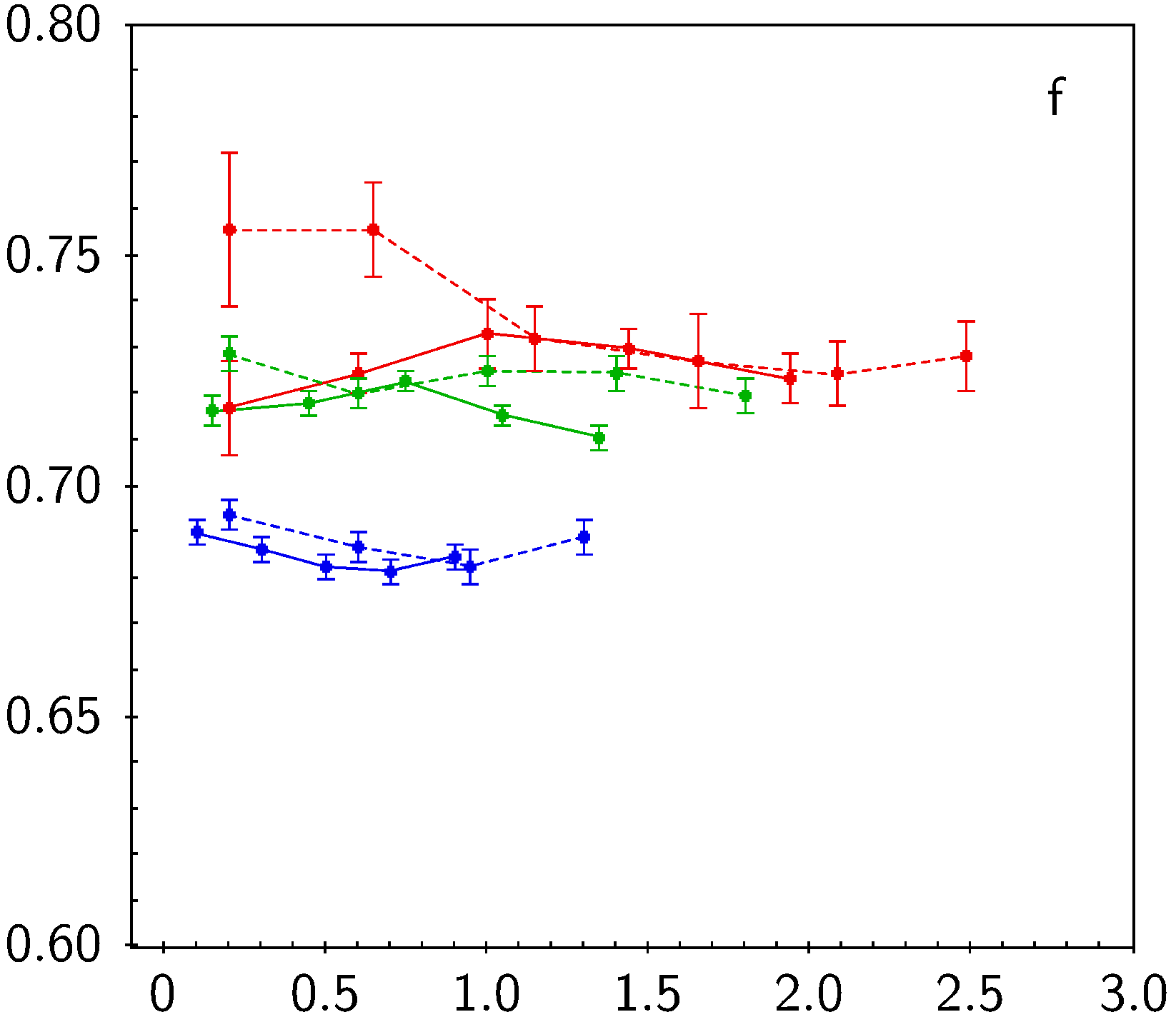}
 \includegraphics[ scale=0.3]{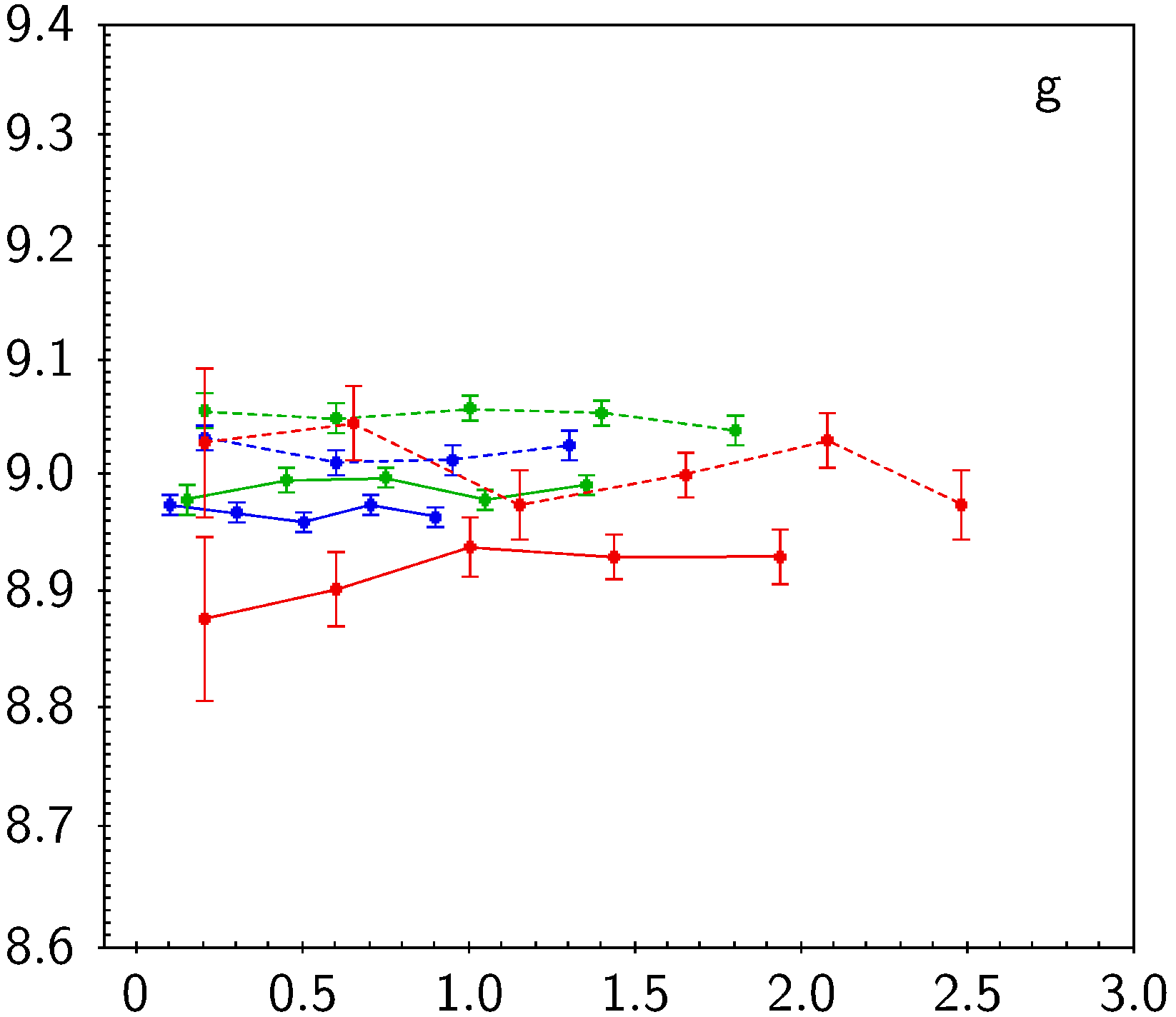}
 \includegraphics[ scale=0.3]{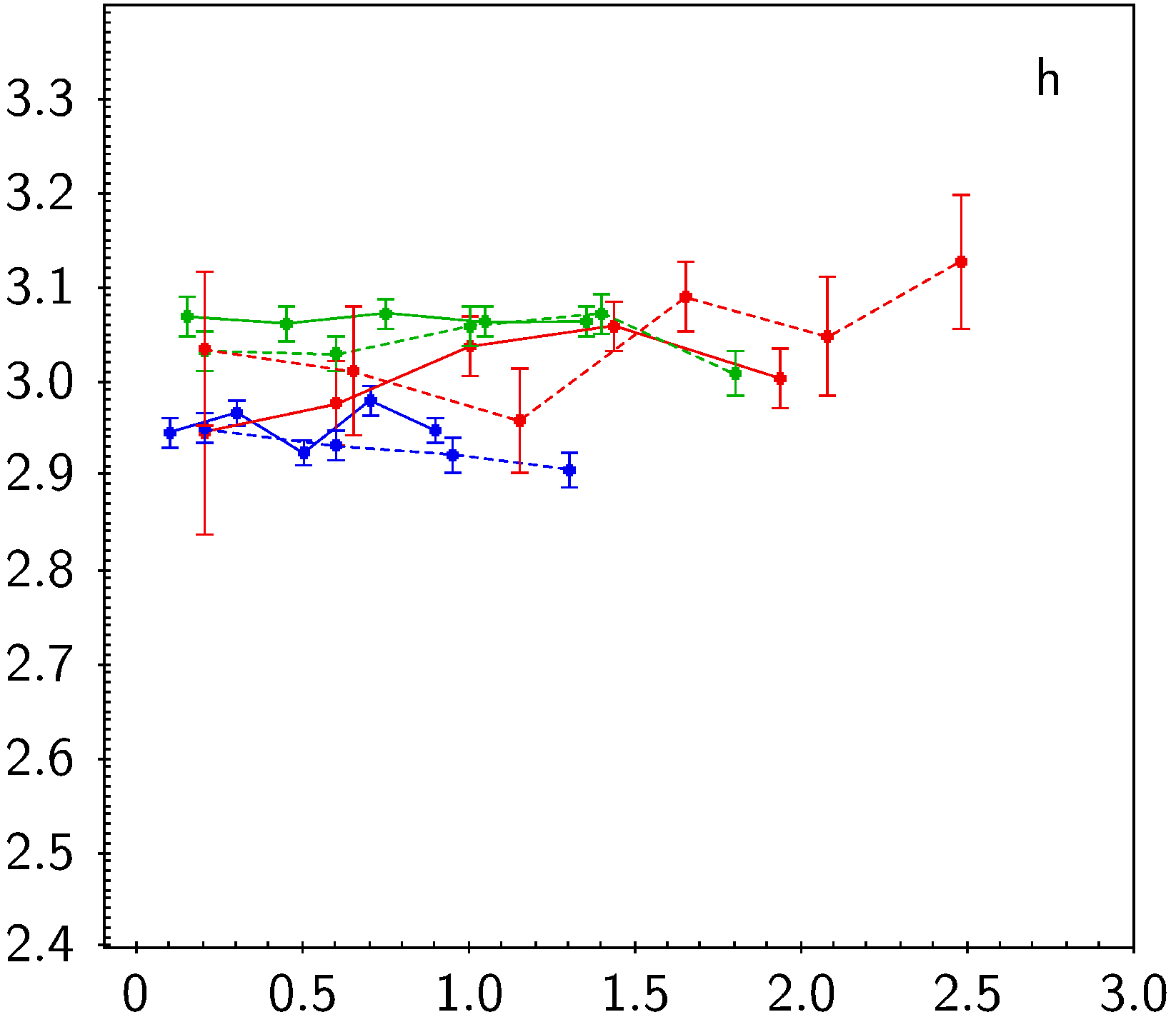}
 \includegraphics[ scale=0.108]{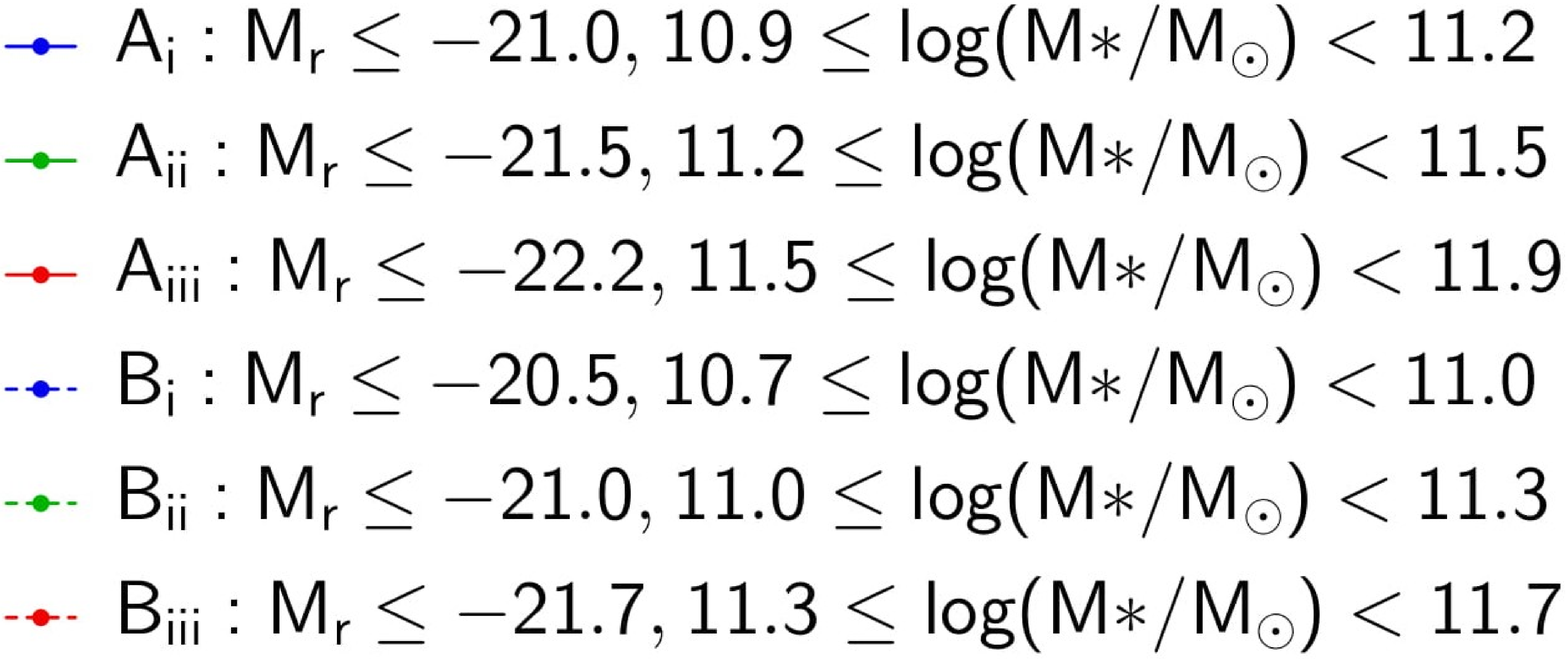}
\figcaption{The optical properties of the BGGs as a function of luminosity gap. 
The y-axis shows the mean of a) M$\rm_{r}$, b) Log[Stellar mass/Solar mass], c) Log[Black hole Mass/Solar mass],
d) Galaxy size/kpc, e) 4000$\AA$ break strength, f) g-r colour, 
g) Log[Surface mass density/Solar mass kpc$^{-1}$] and h) Concentration.
The colours represent different stellar mass ranges and 
BGG absolute magnitude cuts as A-i through to B-iii defined in Section \ref{sec:R}. 
\label{figop}}
 \end{figure*}

\section {Luminosity gap and optical properties of BGGs  }
\label{sec:RO}

\noindent
We first investigate the optical properties of the BGGs for different 
$\Delta$M$\rm_{r}$ systems to study whether there is any characteristic 
that i) changes with $\Delta$M$\rm_{r}$ and so ii) may affect the radio loudness results. 
We also compare our results for the optical properties of BGGs 
against results from a cosmological hydrodynamical simulation.

\subsection {Results of observational data  }
\label{sec:ROO}

\noindent
In Fig.~\ref{figop} we show various optical properties of BGGs as a function 
of luminosity gap for galaxy group sub-samples discussed in Section \ref{sec:R} with N $\geq$ 2.
The y-axis of each plot shows the mean value of each parameter, 
with the error bar given by the standard deviation of the mean.
The solid line and dashed line represent galaxy group
 samples with -20 and -19 cuts respectively. Different 
colours show each sub-samples, i.e., the blue, green and red
lines represent low mass, medium mass and massive galaxies respectively. 
We note that we only display luminosity gap
bins which preserve the luminosity gap completeness 
and we applied the upper limit $\Delta$M$\rm_{r}$ cut
to all sub-samples.
The mean of the stellar mass and the r-band magnitude confirm that the sample is 
unbiased for each bin of magnitude gap. There are no significant changes in any of the other optical
properties with respect to the $\Delta$M$\rm_{r}$. 
The results for N$\geq$ 3, 4 and 5 are not shown but in all cases they confirm the N$\geq$2 results. 

These results show that, once selected at constant stellar mass, 
the optical properties of the BGG within a group do not display 
any dependencies upon the luminosity gap. This is consistent with the results of 
Trevisian et al. (2017), who find no correlation between the magnitude gap of 
the group and the optical properties of the BGG such as age, 
metallicity, [$\alpha$/Fe] and star formation history.

\subsection{Results of hydrodynamical simulation}
\label{sec:Simu}

\begin{figure*}
\centering
 \includegraphics[ scale=0.22]{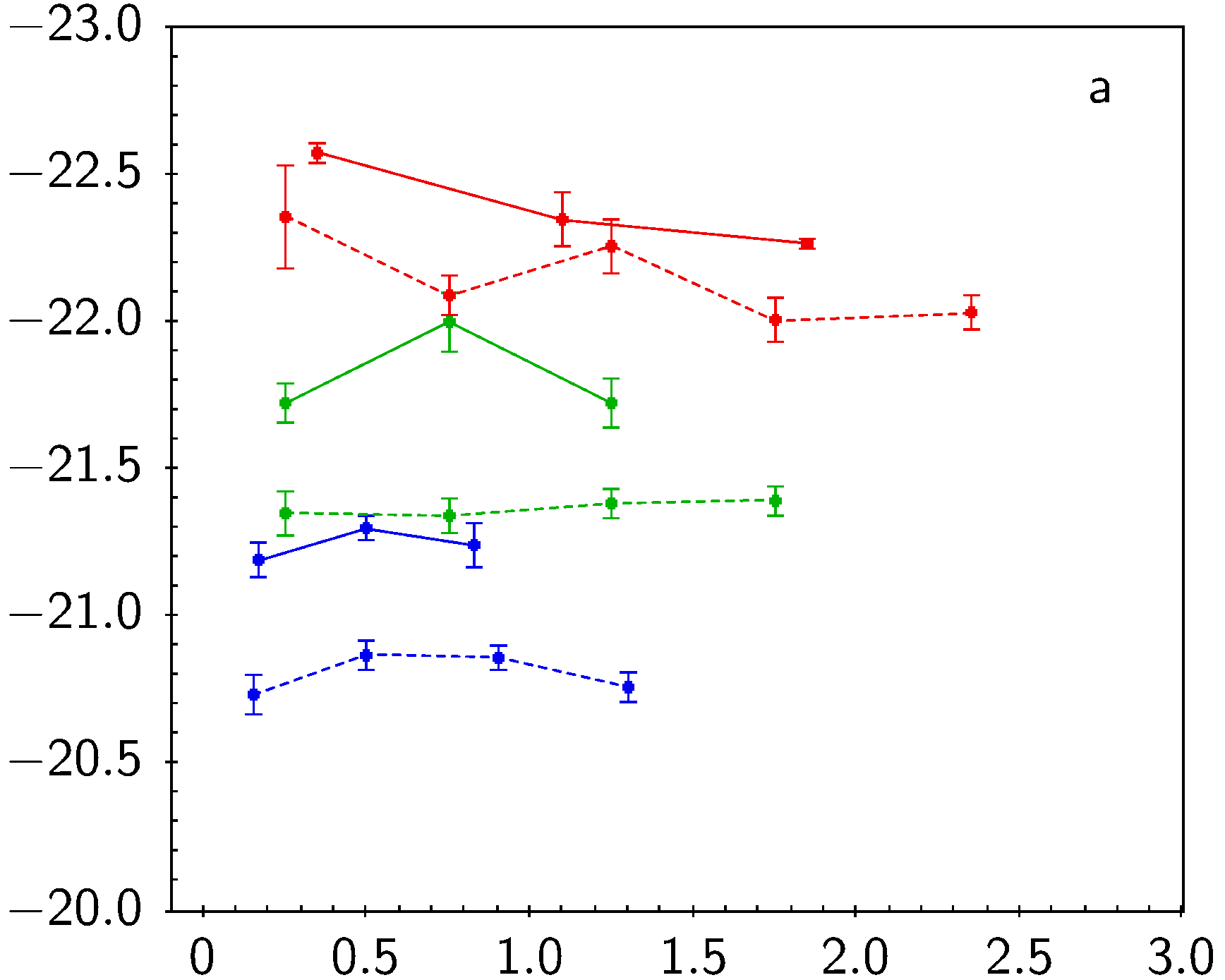}
 \includegraphics[ scale=0.22]{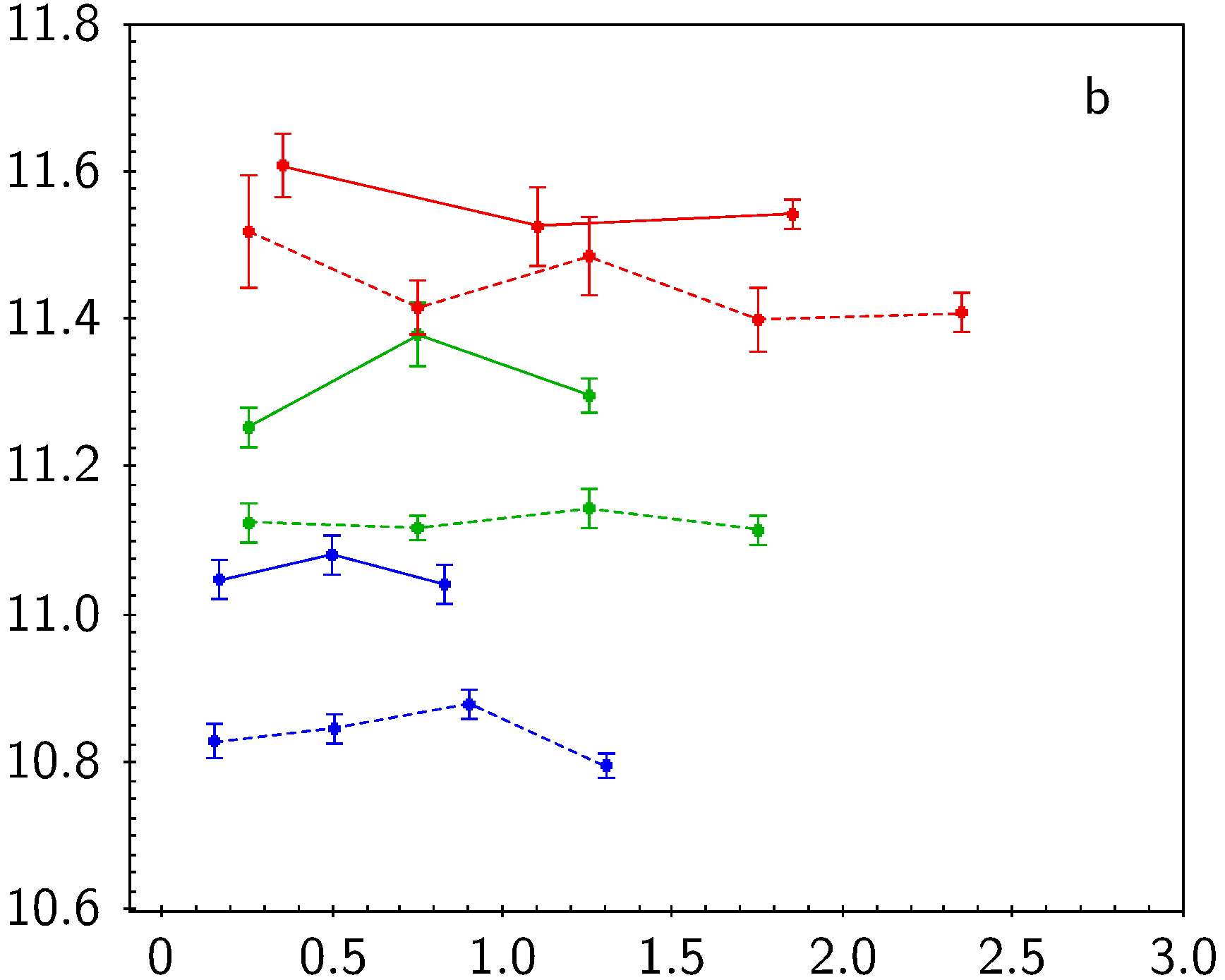}
 \includegraphics[ scale=0.22]{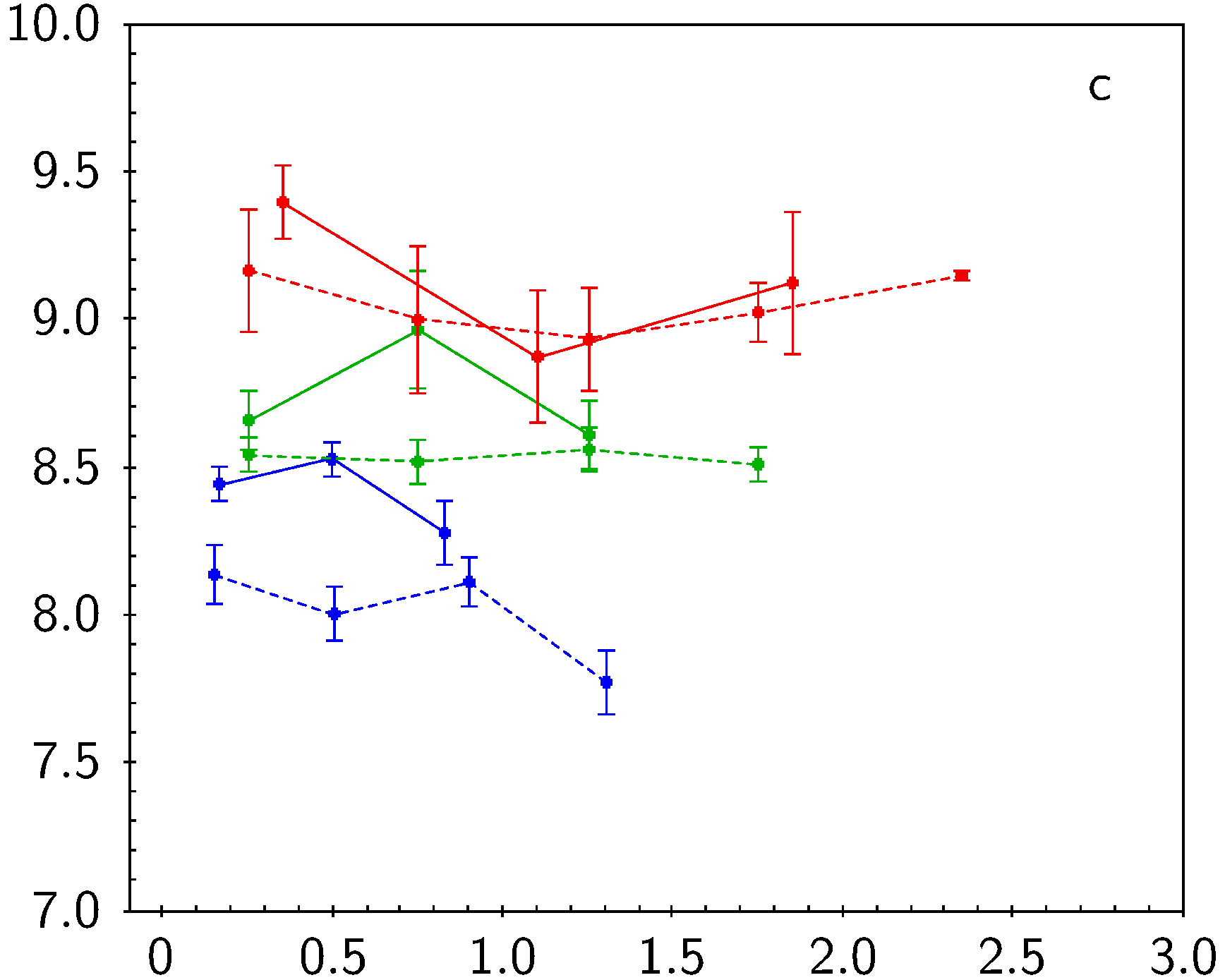}
 \includegraphics[ scale=0.22]{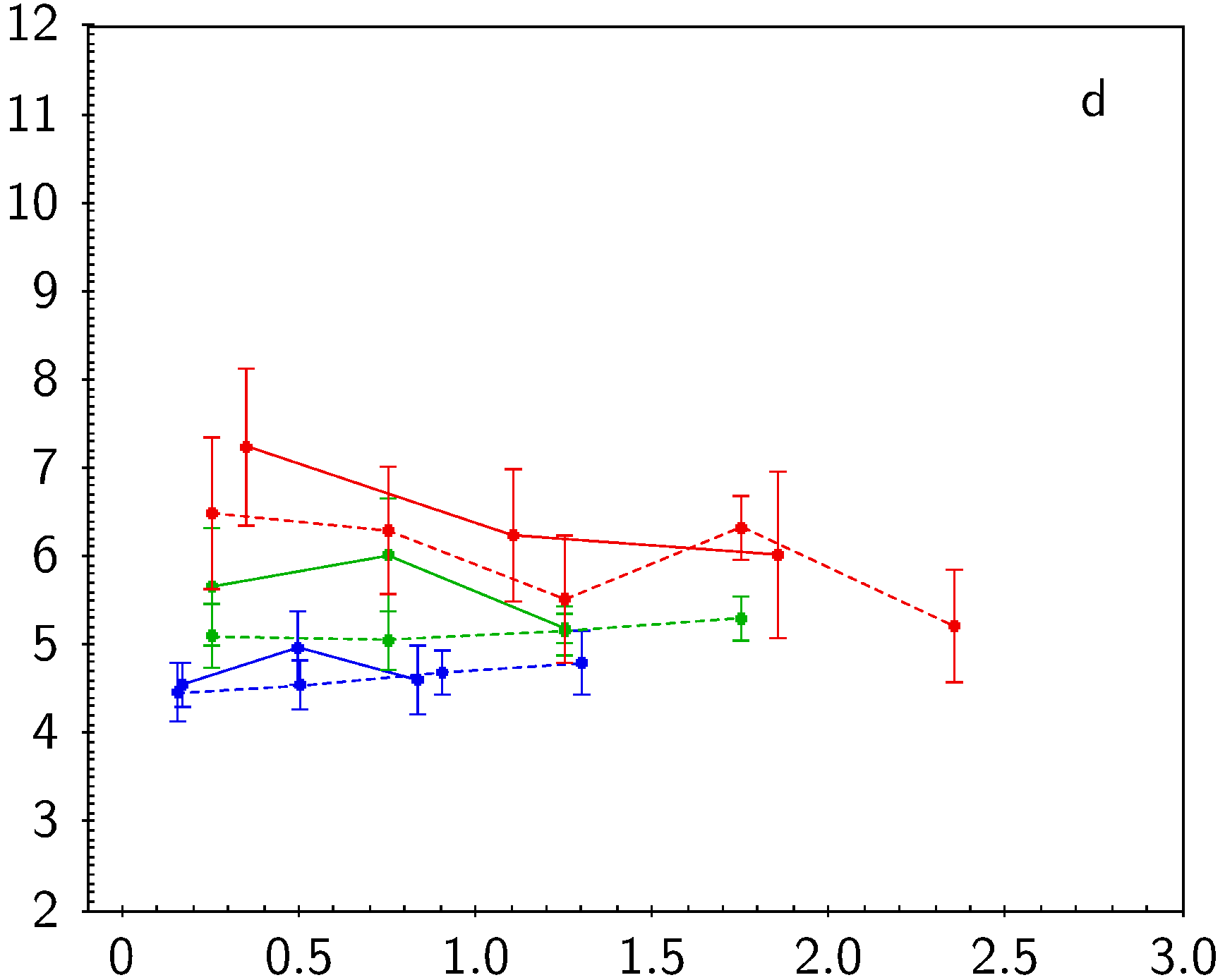}
 \includegraphics[ scale=0.22]{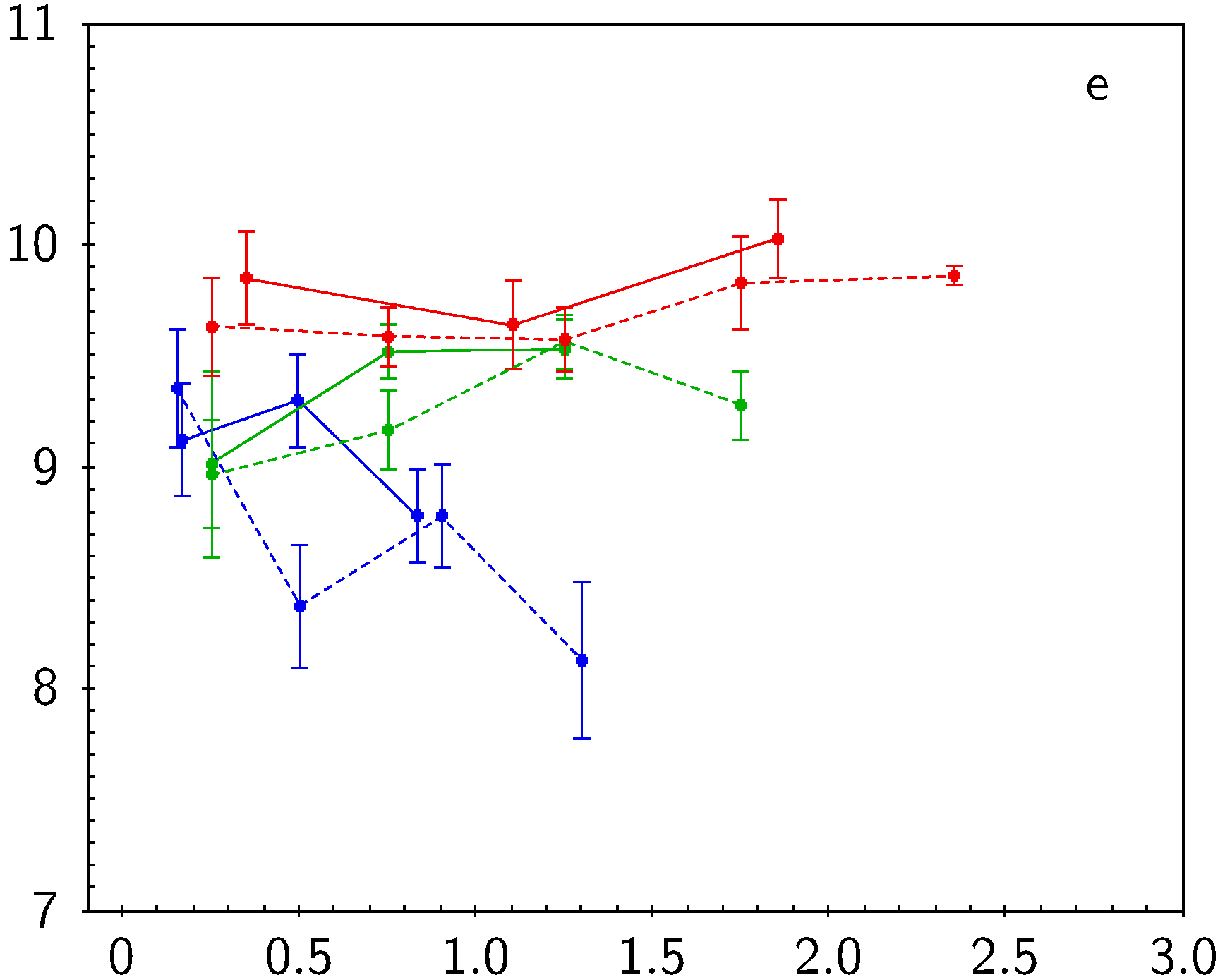}
 \includegraphics[ scale=0.22]{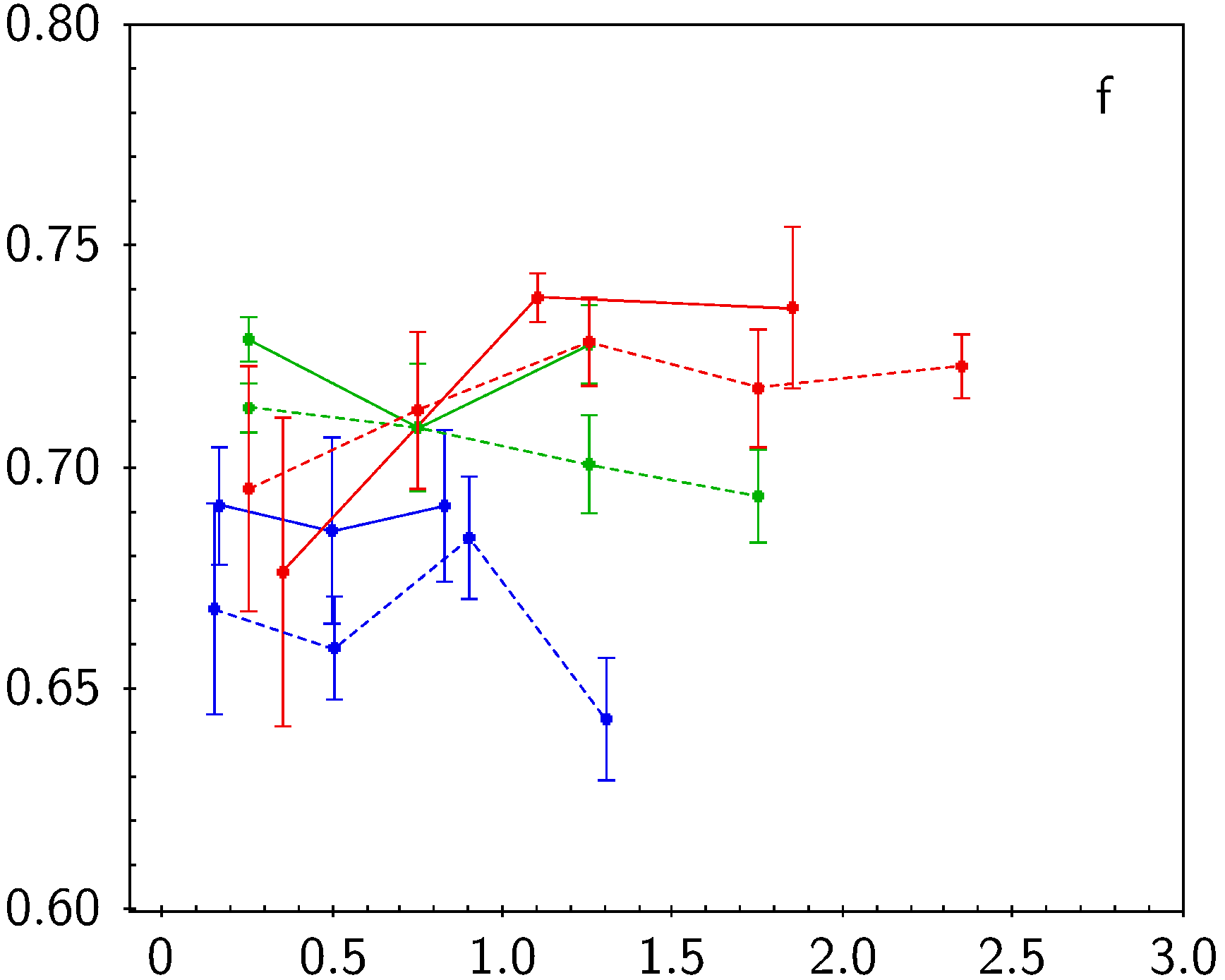}
 \includegraphics[ scale=0.22]{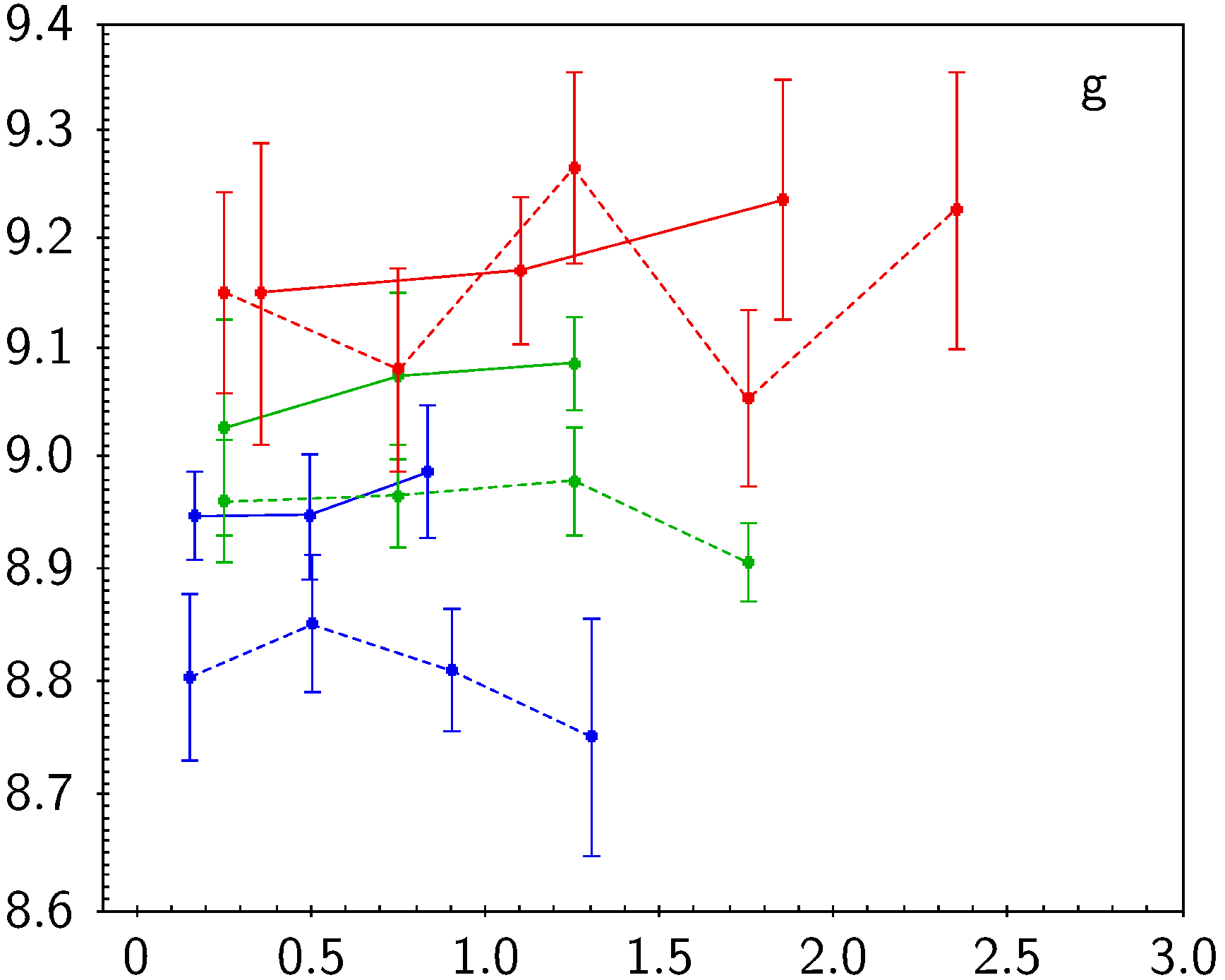}
 \includegraphics[ scale=0.082]{legend3-4.eps}
\figcaption{The properties of the BGGs drawn from the EAGLE hydrodynamical 
simulation as a function of group luminosity gap. The y-axis is 
the mean of a) M$\rm_{r}$, b) Log[Stellar mass/Solar mass], c) Log[Black hole Mass/Solar mass],
d) Galaxy size/kpc, e) Stellar age/Gyr, f) g-r colour and 
g) Log[Surface mass density/Solar mass kpc$^{-1}$]. We replicated the same sample 
selection criteria for the simulation data
 as were described for the observational data in  Section \ref{sec:R} and plotted in Fig.~\ref{figop}; 
different colours and lines represent 
 different sub-samples, as in Fig.~\ref{figop}. Plots with the same labels as in Fig.~\ref{figop}
present the same properties to compare.
The results agree with the observations and show lack of any
trend between properties of the BGGs and the luminosity gap.
\label{fighy1}}
 \end{figure*}

\begin{figure*}
\centering
 \includegraphics[ scale=0.22]{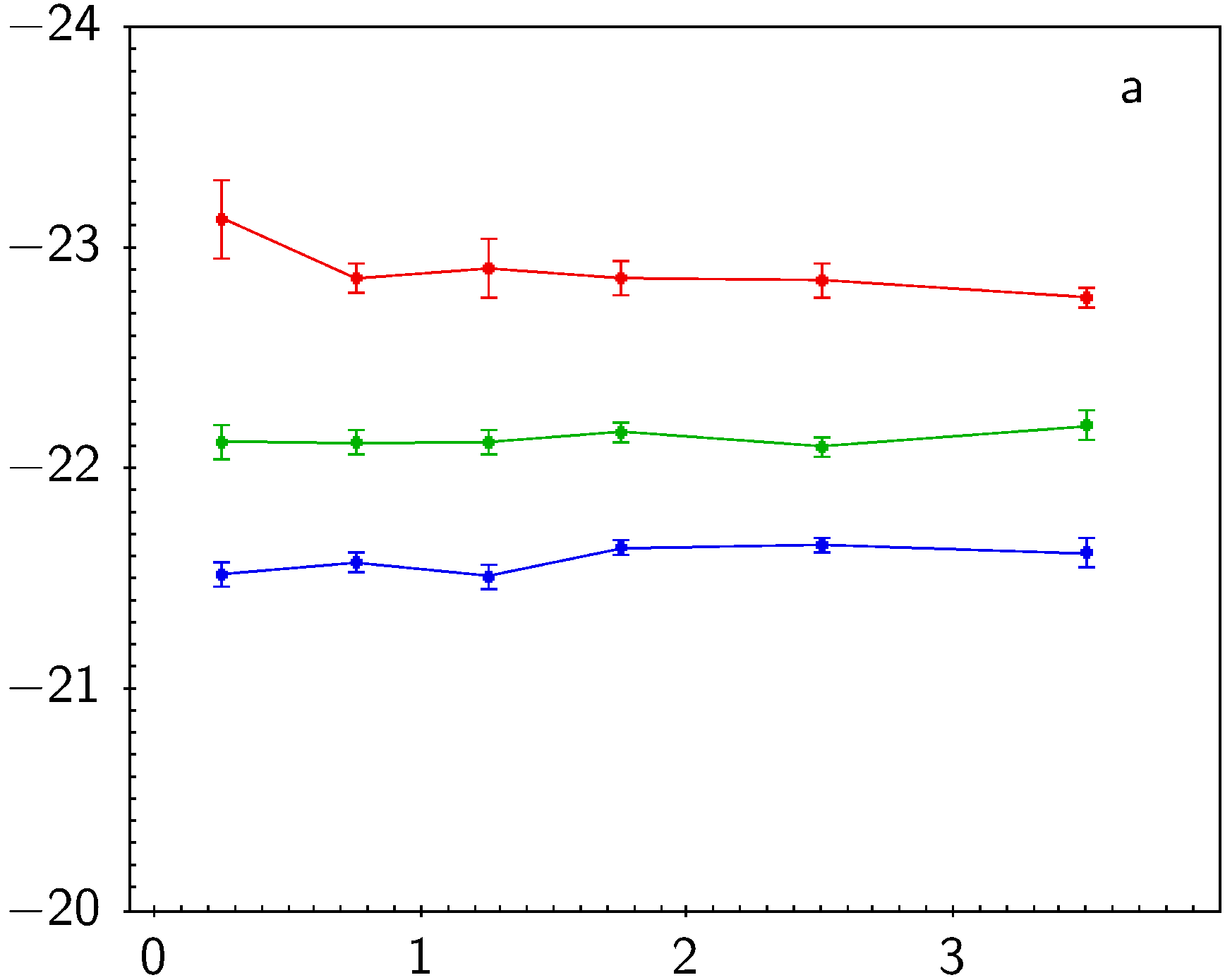}
 \includegraphics[ scale=0.22]{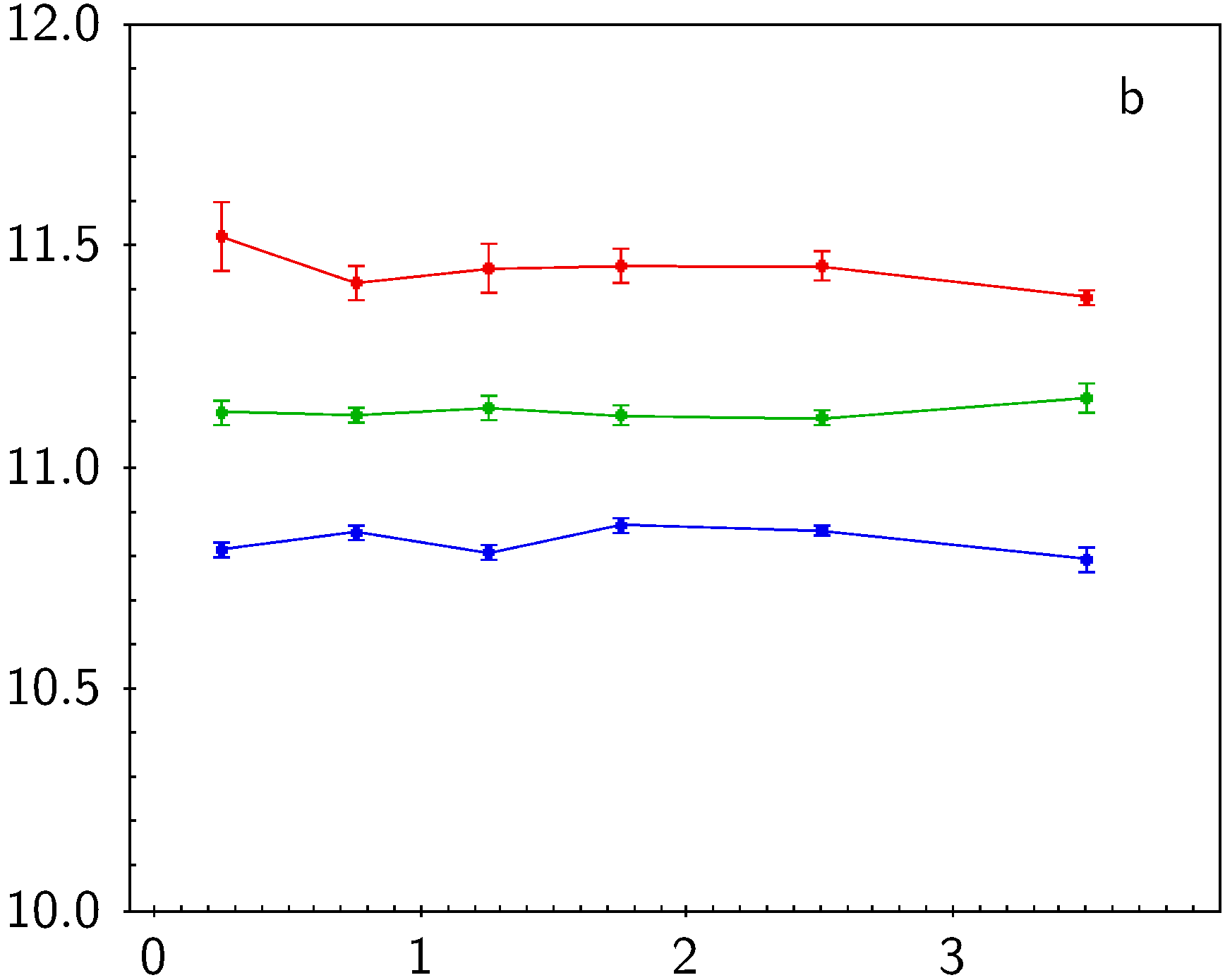}
 \includegraphics[ scale=0.22]{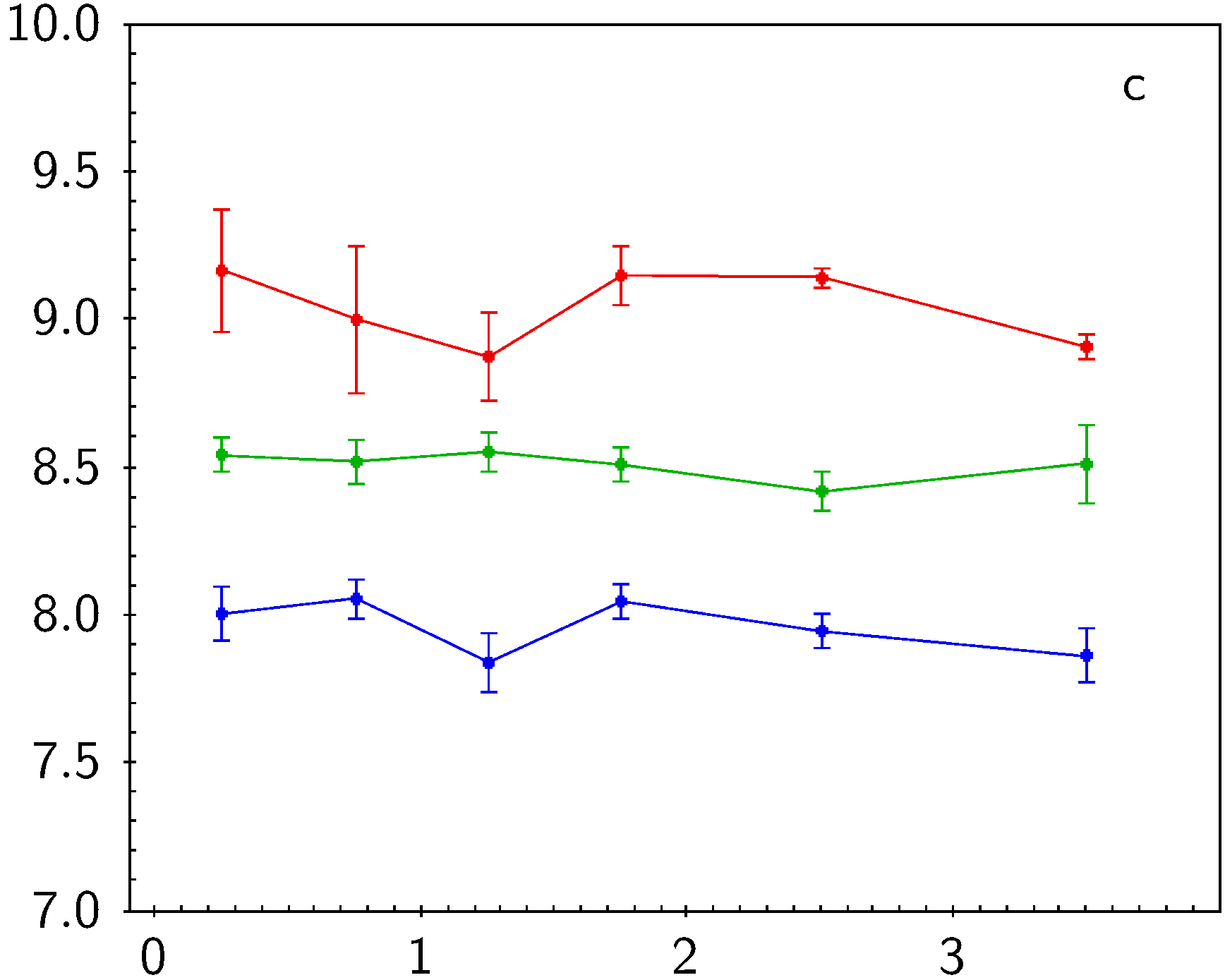}
 \includegraphics[ scale=0.22]{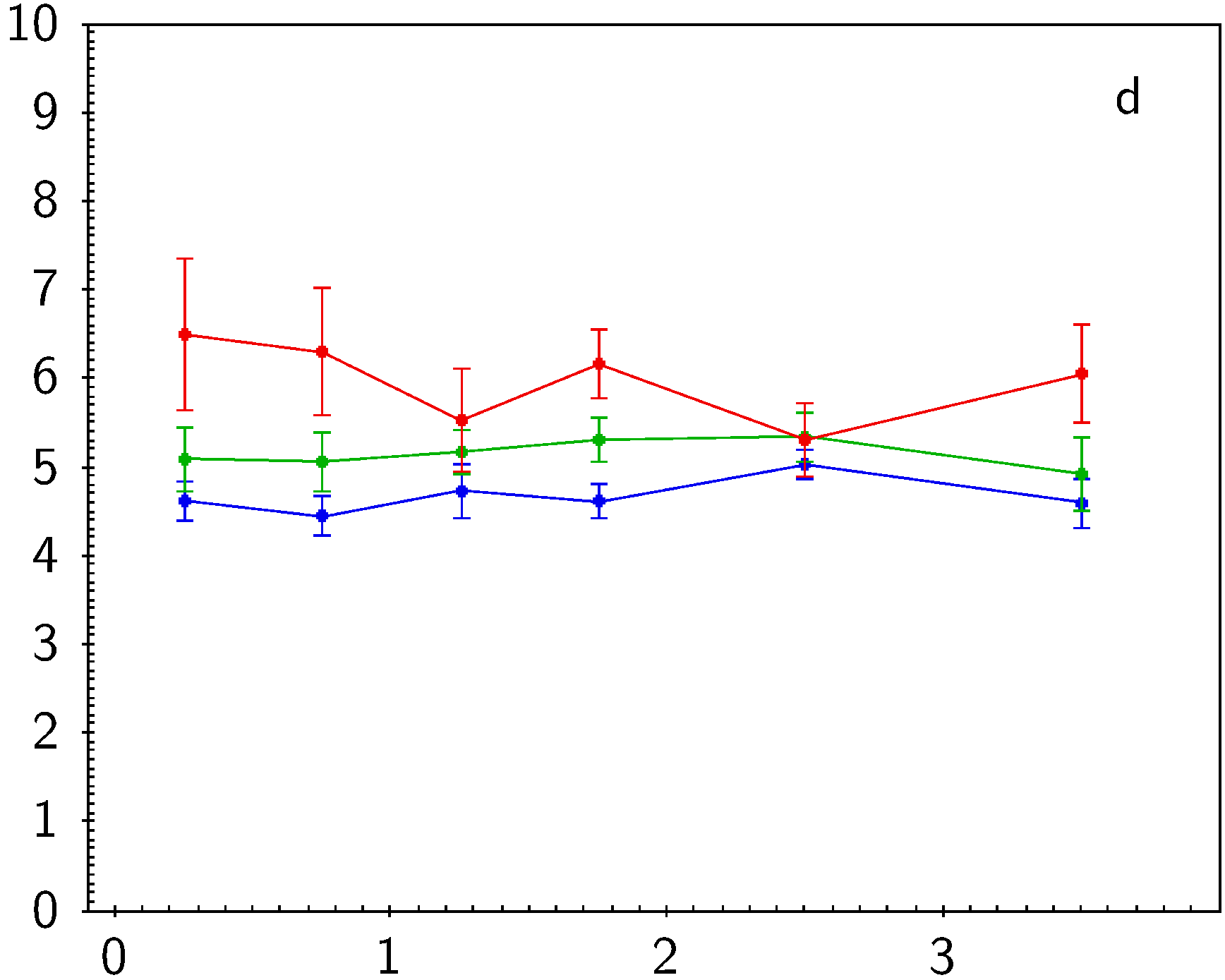}
 \includegraphics[ scale=0.22]{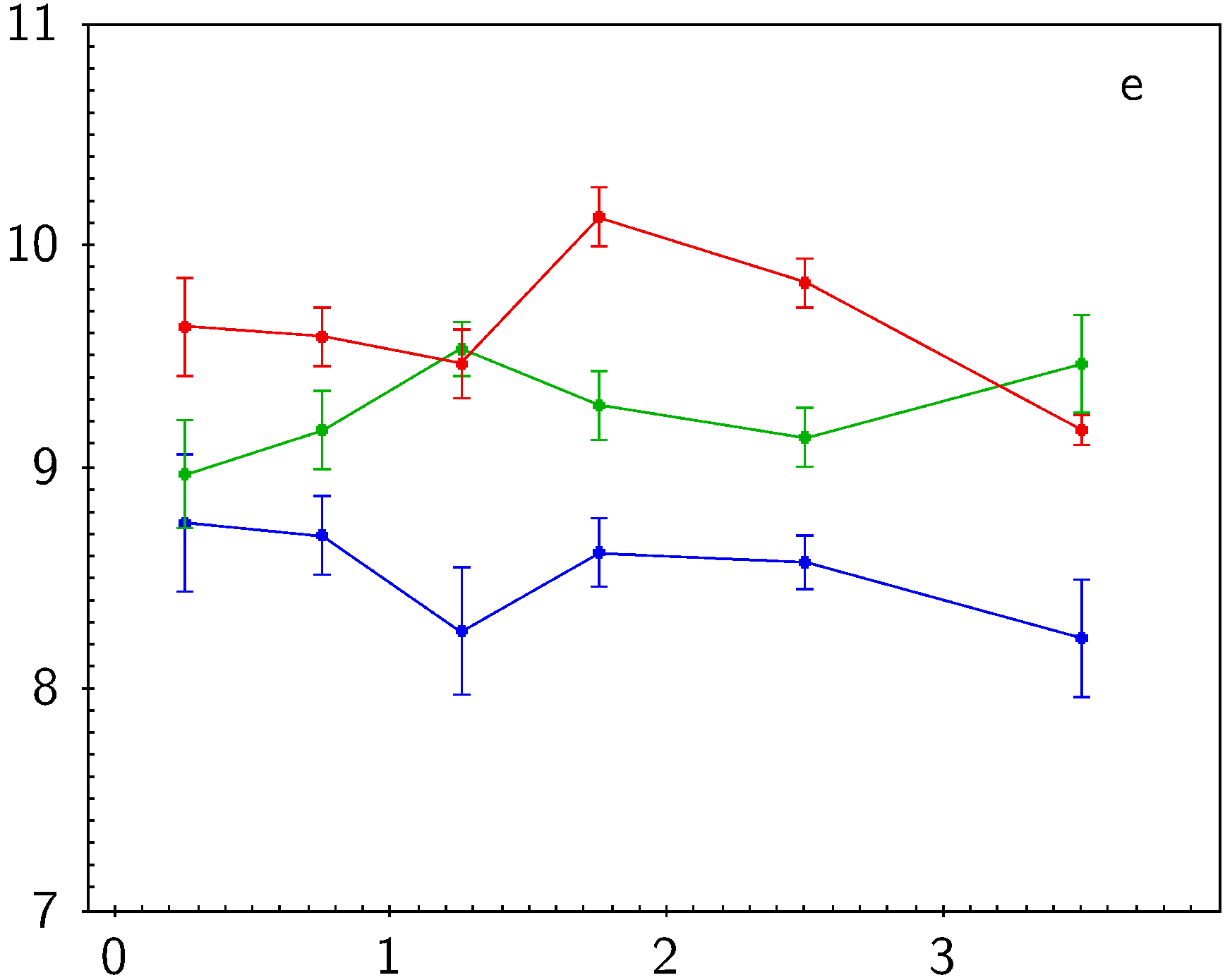}
 \includegraphics[ scale=0.22]{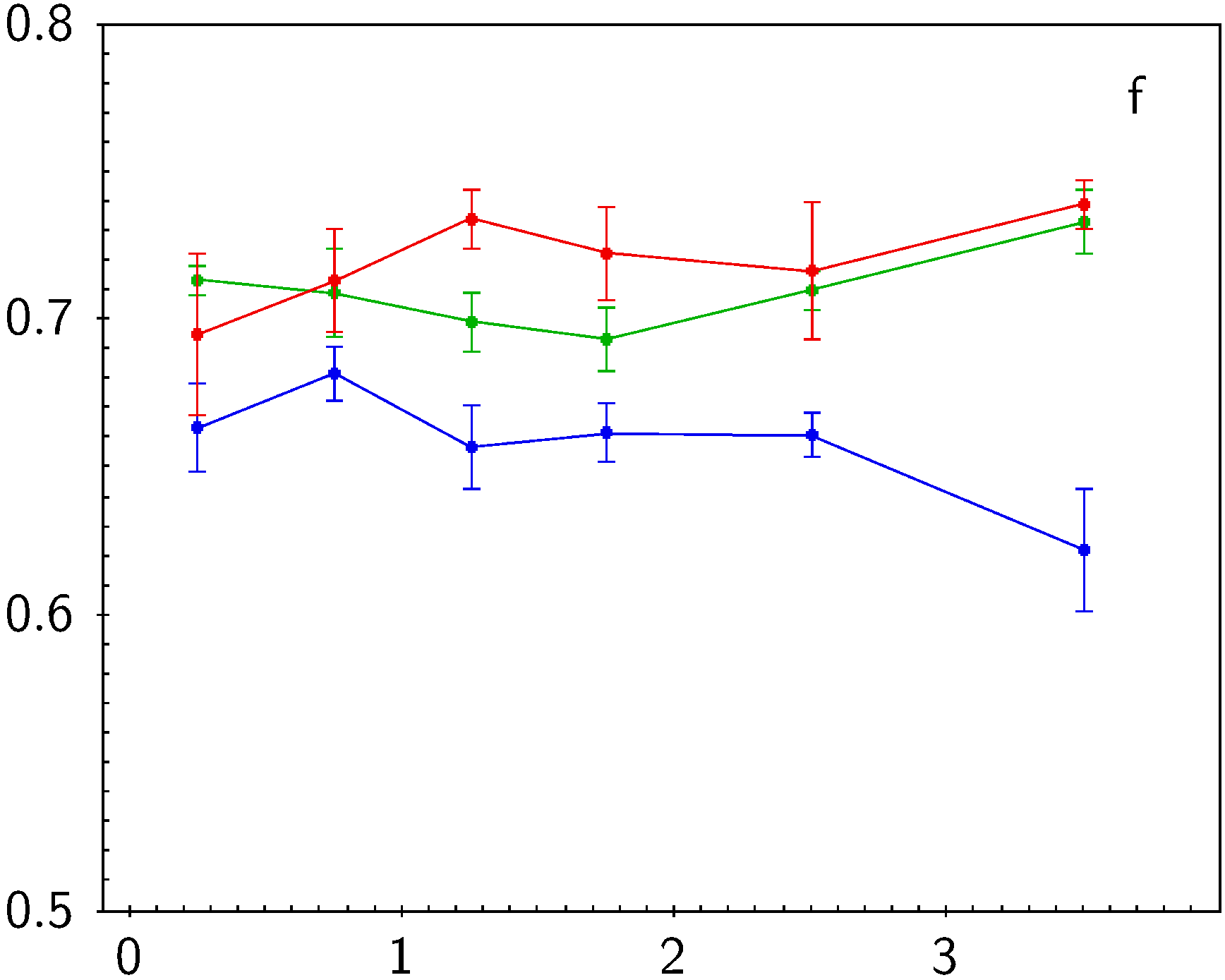}
 \includegraphics[ scale=0.22]{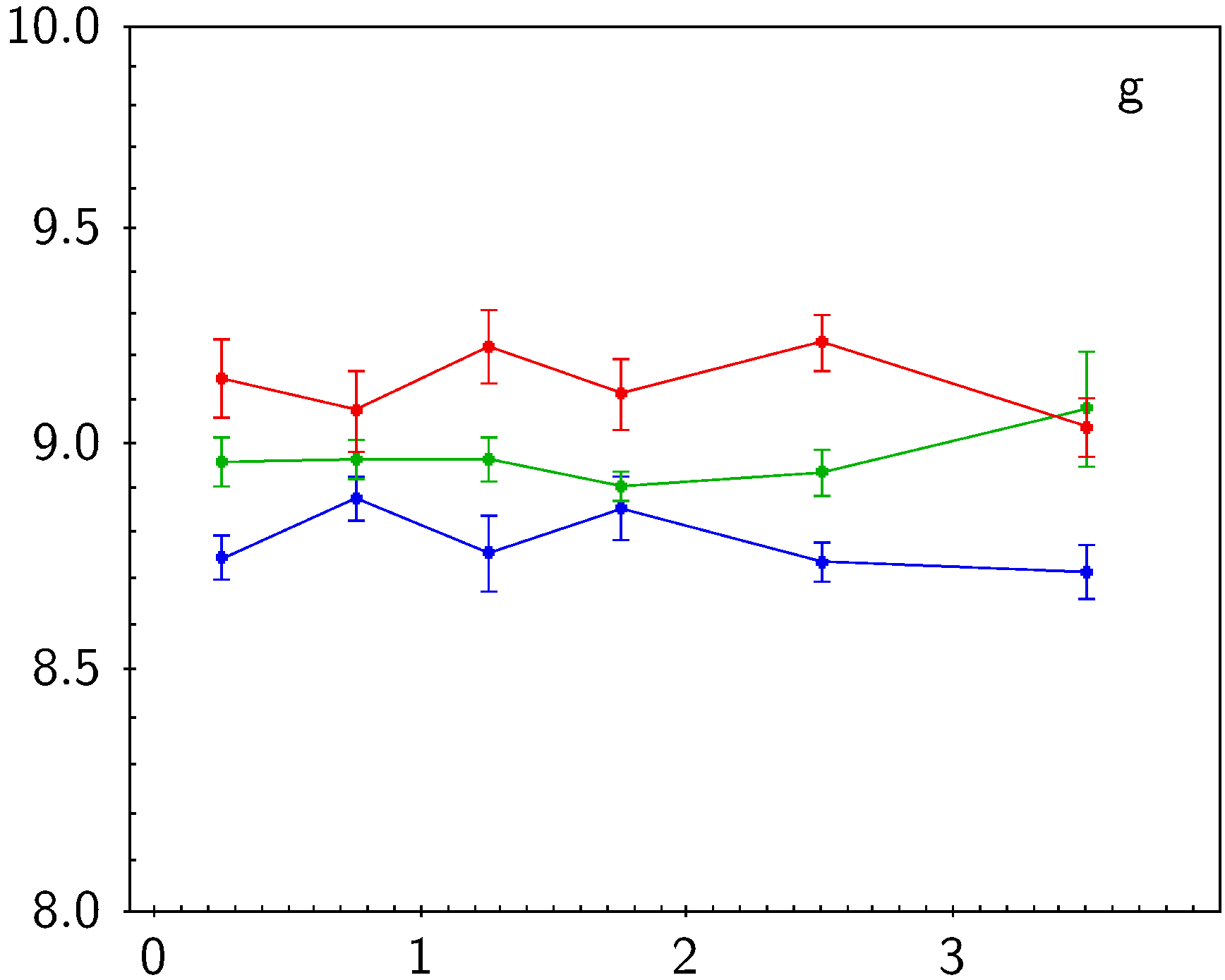}
 \includegraphics[ scale=0.126]{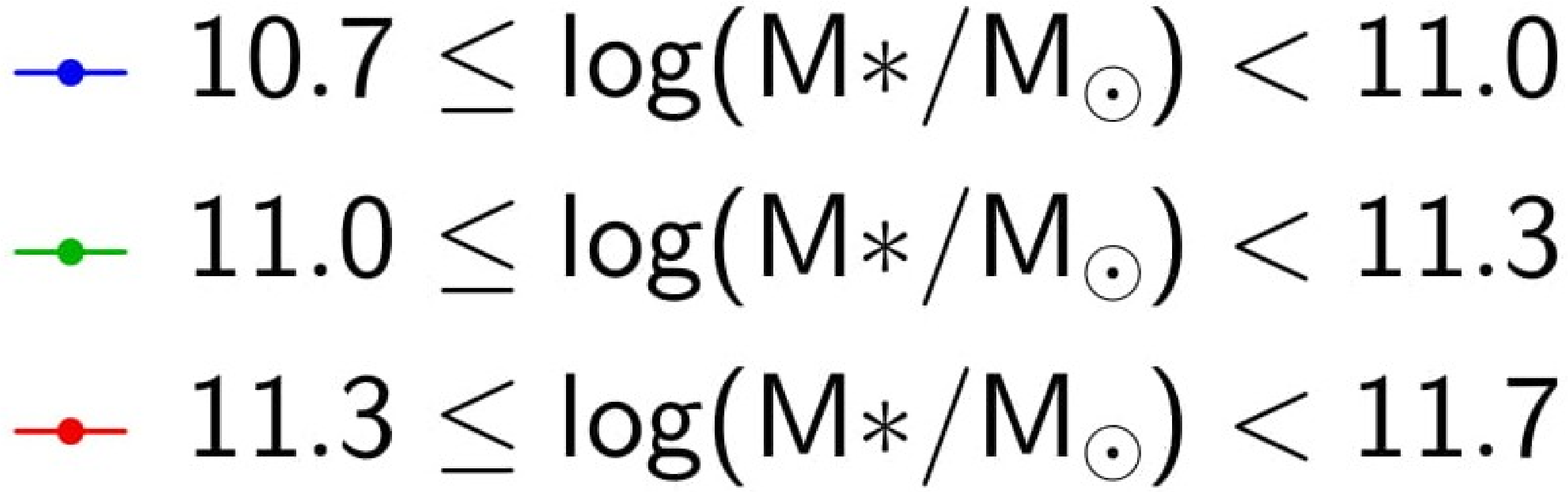}
\figcaption{As Fig.~\ref{fighy1}, but with no restriction in magnitude applied in the 
construction of the group catalogues from the simulation data, in order to extend analysis to higher luminosity gaps.
The x-axis represents $\Delta$M$\rm_{r}$. The y-axis is 
the mean of a) M$\rm_{r}$, b) Log[Stellar mass/Solar mass], c) Log[Black hole Mass/Solar mass],
d) Galaxy size/kpc, e) Stellar age/Gyr, f) g-r colour and 
g) Log[Surface mass density/Solar mass kpc$^{-1}$].
The colours represent different mass bins as defined in Fig.~\ref{figop} for -19 cut sample. 
Plots with the same labels as in Fig.~\ref{figop}
present the same properties to compare.
The optical 
properties of the BGGs show no trends with the group luminosity gap.
\label{fighy2}}
 \end{figure*}

\noindent
In order to compare our optical results, and to ensure that it is not
the case that luminosity-gap dependencies only begin to appear at 
larger luminosity gaps than our observational sample is able to probe, we make use of the EAGLE
simulation.
EAGLE is a cosmological hydrodynamical simulation which tracks the evolution
of baryonic and nonbaryonic matter from z=127 to z=0 (Schaye et~al.\ 2015, Crain et~al.\ 2015).
McAlpine et~al.\ (2016) provide a database for querying the halo and galaxy catalogues
including FOF group members extracted from the simulation. We used version Ref-L100N1504 of 
the data release and snapshot 28, corresponding to z=0. 
The parameters from the simulation analagous to our observational 
data are as follows: the stellar mass, black hole mass, stellar age, 
projected half mass radius (projected physical radius enclosing half of 
the stellar mass averaged over three orthogonal projections),
 g-band and r-band magnitudes. The g-r colour is calculated from the g-band and r-band magnitudes 
with dust modelling presented in Trayford et~al.\ (2015).
We used a 30 pkpc (proper distance) aperture size for extracting the parameters to allow proper 
comparison with observations which extract physical parameters from aperture-based measurements. 
The data were extracted from Table B1-B5 of McAlpine et~al.\ (2016) 
using the SQL query presented in the Appendix A of this paper.

To investigate BGG properties, we extract all z=0 galaxies with stellar masses greater 
than 10$^{9}$~\msun, since EAGLE galaxy properties become 
unreliable below this stellar mass (McAlpine et~al.\ 2016;
Schaye et~al.\ 2015). The sample
includes 6593 galaxies with 858 unique galaxy groups of 2 or more galaxy members.
To make a direct comparison with the observational data, we applied the 
same r-band absolute magnitude cuts as we applied to the observational 
data in Section \ref{sec:R}, namely -20 (or -19) r-band absolute magnitude 
cuts for group member definition, and -21.0 (20.5), -21.5 (-21.0) 
and -22.2 (-21.7) cuts for the BGGs, as well as the corresponding 
stellar mass cuts outlined in Section \ref{sec:R}. This reduces the 
sample size but allows us to compare the simulation directly against 
the observational results. Our results are plotted in Fig. \ref{fighy1}.  
All of the results agree well with the observations (Fig. \ref{figop}) 
both in the lack of trend and in the normalisation. This gives us 
confidence in our observational samples.

Next, we investigate whether our observational limitations are 
influencing our ability to detect any trends. To do this, we return 
to the 858 unique galaxy groups and instead apply no absolute magnitude 
cuts but only stellar mass cuts (defined to  match those of the 
M$\rm_{r}$$<$-19 sub-samples). This larger EAGLE sample allows us 
to explore the luminosity gap up to higher $\Delta$M$\rm_{r}$ 
(up to $\Delta$M$\rm_{r}$$\sim$4). The results are displayed in 
Figure \ref{fighy2}. This shows that the flat trends continue 
to the highest $\Delta$M$\rm_{r}$ probed in the simulations 
(although normalisations change due to the removal of the 
absolute magnitude limit), and that the lack of any trend in 
the observed data is thus not due to the limited parameter space of our observations.

\section {Luminosity gap and radio properties of BGGs  }
\label{sec:RR}

\begin{figure*}
\centering
 \includegraphics[scale=0.45]{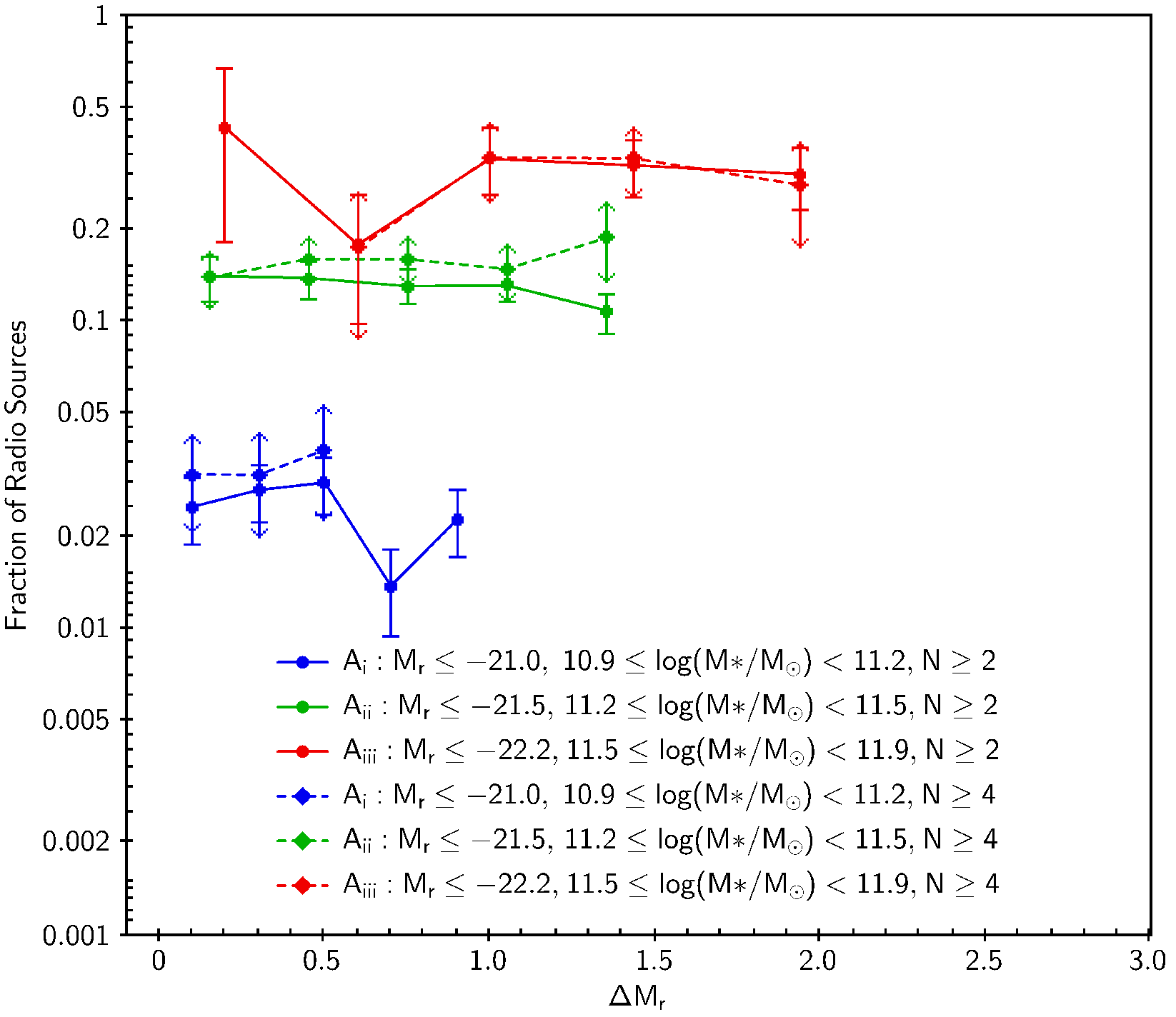}
 \includegraphics[scale=0.45]{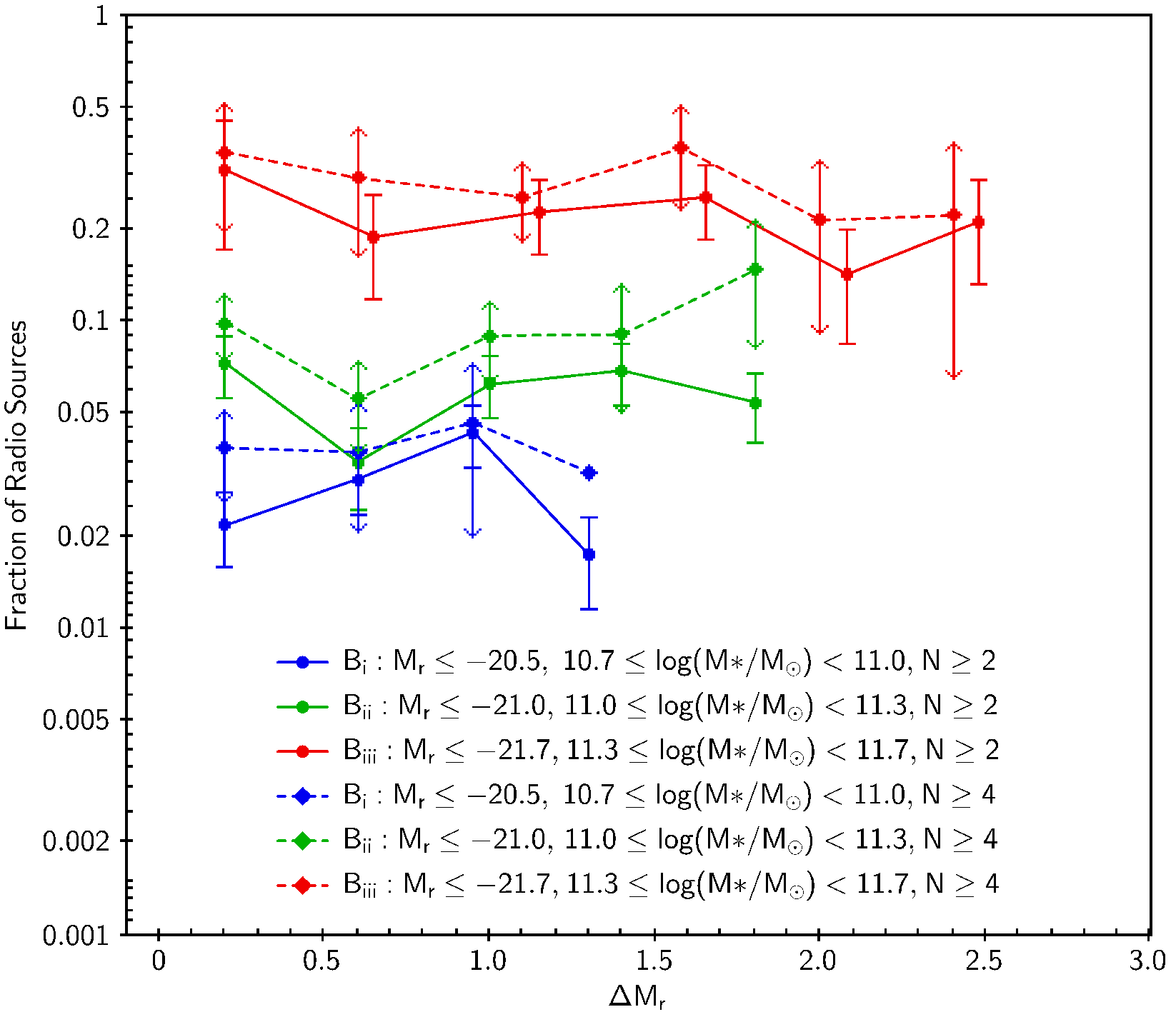}
\figcaption{The radio-loud fraction ($\it{n}$/$\mathcal{N}$) as a function of luminosity gap 
where $\it{n}$ is the number of BGGs hosting radio-loud AGN with radio 
luminosities above the selected luminosity limit, and $\mathcal{N}$ is the total number of BGGs in each bin.
 The error bars (capped-lines and arrows) are  $\sqrt{\it{n}}$/$\mathcal{N}$.
The left (right)-hand panel shows the results for the group 
catalogue selected above r-band absolute magnitude -20 (-19), 
with a radio luminosity limit of 10$^{22.9}$ (10$^{22.6}$) W Hz$^{-1}$ (see Fig. \ref{figsample}). 
Solid (dashed) lines show groups with richness N$\geq$2 (N$\geq$4).
 The colours represent different stellar mass ranges and magnitude cuts as in Fig.~\ref{figop} 
and defined in Section \ref{sec:R}.  
\label{figradio}}
 \end{figure*}

\noindent
The work in the previous section confirmed that there is no correlation 
between the optical properties of the BGGs and the luminosity gap in 
the subsamples that we have defined, which could drive a secondary 
dependence of the radio properties on luminosity gap. The next step 
is then to investigate the radio properties of the BGGs.
We calculate the fraction of radio loud BGGs in each bin of luminosity gap for
each galaxy group samples in the three different mass ranges. 
The 3mJy flux density limit of the NVSS
has been used to calculate the detection limit for the samples with -19 and -20 cuts at the
redshifts of 0.07 and 0.11 respectively. This  
corresponds to a 1.4 GHz luminosity of 3$\times$10$^{22}$ W Hz$^{-1}$ for -19 cut and  
9$\times$10$^{22}$ W Hz$^{-1}$ for -20 cut samples. Therefore, the radio samples are complete
to these limits up to the highest redshifts for each.

The \textquoteleft radio-loud fraction\textquoteright~  
is simply the fraction of all the BGGs in a given bin of luminosity gap 
and mass which satisfy two conditions: (a)~they have a radio luminosity 
greater than the completeness limit; and (b)~they are labelled as 
radio-AGN based on the AGN/SFG separation in Best \& Heckman (2012). 
We note that the Best \& Heckman (2012) catalogue also includes the excitation index of the radio AGN, with each source classified as either a high excitation radio galaxy (HERG) or a low excitation radio galaxy (LERG).  The BGG sample in this study is dominated by sources classified as LERGs with only few sources classified as HERGs ($<$1$\%$), thus the results represent low-accretion-rate-mode AGN activity.
The radio-loud fractions are illustrated in Fig.~\ref{figradio}.
The solid and dashed lines show the results for galaxy groups
with richness N $\geq$ 2 (circle) and N $\geq$ 4 (diamond) respectively. 
Different colours are defined as in Fig.~\ref{figop}. 
 The error bars are calculated 
using a Poissonian approach. There are no differences in 
the radio-loud fraction for different luminosity gap bins
for any of the samples A-i to B-iii. 
The results for richness 2 and 4 are consistent with 
each other especially for massive BGGs where 
we have the largest magnitude gap bin. This shows that the results are not contaminated by the
group finding uncertainties. We also note that the same results are obtained if we change 
the radio luminosity condition (a) to an order of magnitude higher value to select 
only very luminous radio AGN as the radio-loud AGN. Therefore, the result is independent of the radio
luminosity limit we applied.

From the results presented in Fig.~\ref{figradio}, we conclude that  
the radio loudness properties of the BGGs
are much more strongly correlated with stellar mass than with 
 luminosity gap, and, by implication, local environment, at least 
up to the magnitude gaps that we can investigate with the current sample.

\begin{figure*}
\centering
 \includegraphics[ scale=0.45]{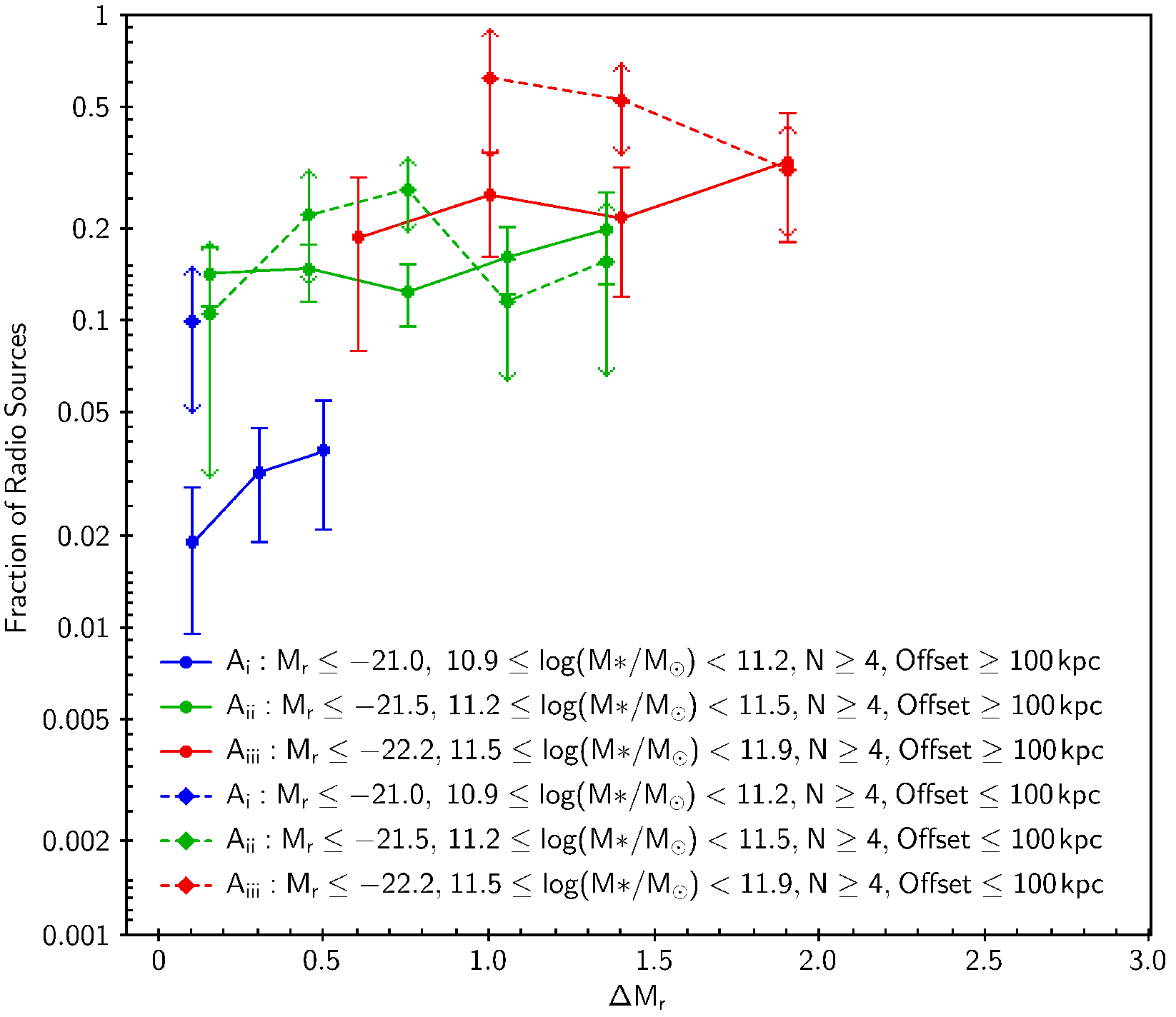}
 \includegraphics[ scale=0.45]{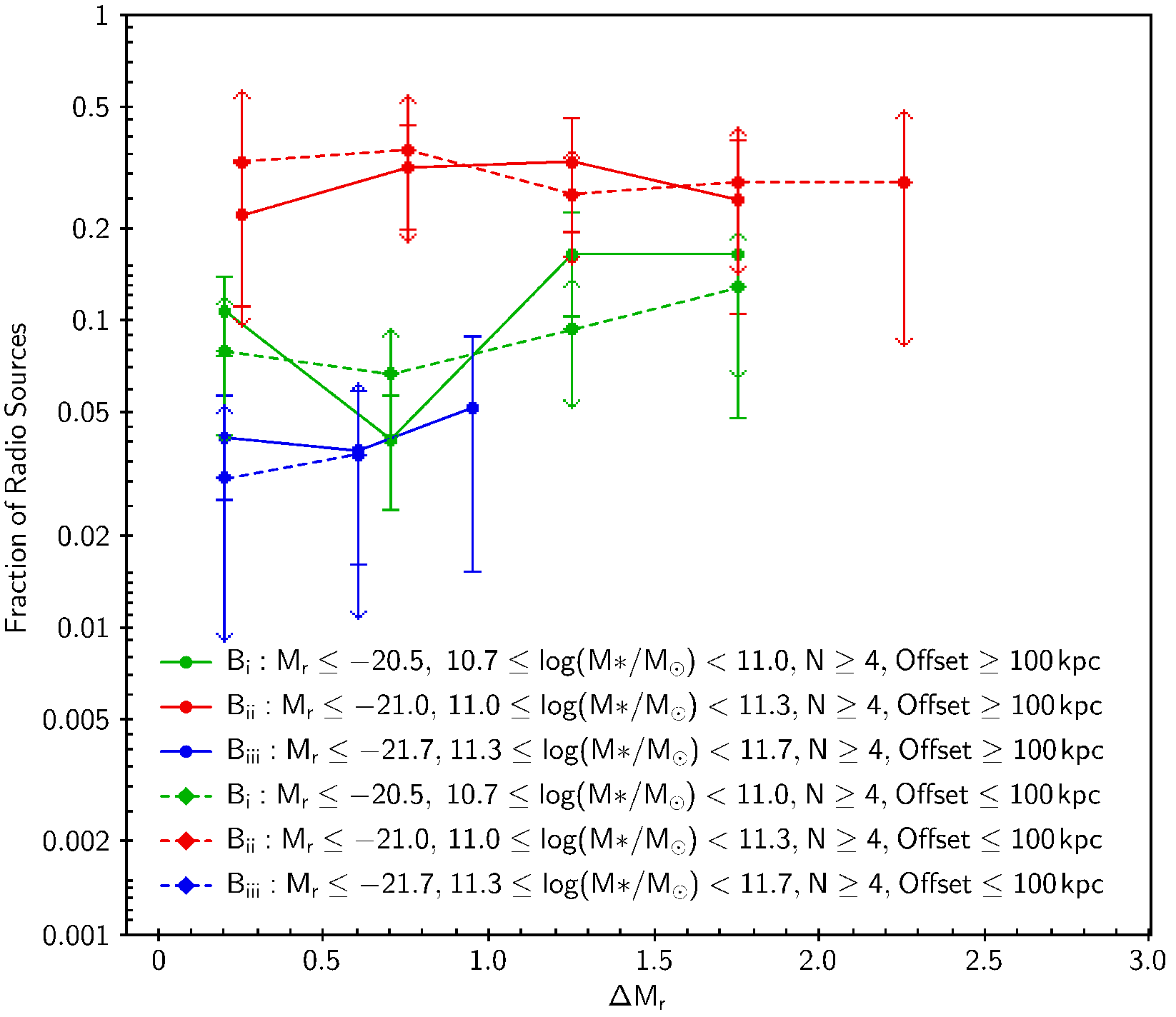}

\figcaption{The radio-loud fraction with radio luminosities above 
10$^{22.9}$ W Hz$^{-1}$ (left-hand panel) and 10$^{22.6}$ W Hz$^{-1}$ (right-hand panel)
for -20 and -19 cut samples (see Fig. \ref{figsample}) respectively, as a function of luminosity gap,
for galaxy groups with richness of N$\geq$4. The fraction and error bars are defined as in Fig.~\ref{figradio}.
The solid line represents BGGs outside the 100 kpc radius  
from the center of group and the dashed line represent those inside the 100 kpc radius 
from the center of group. The colours 
represent different stellar mass ranges and magnitude cuts as in Fig.~\ref{figop} 
and defined in Section \ref{sec:R}.
\label{figoff}}
 \end{figure*}

\section{BGG Offset from the center of groups}
\label{sec:off}

\noindent
In this Section, we investigate the impact of BGG offset from 
the center of the group (hereafter the offset) on the fraction of radio 
AGN. Raouf et~al.\ (2014; 2016) showed that BGG offset can be used 
as a proxy for halo relaxation; a system with a 
large offset is regarded as an unrelaxed system.  
A galaxy group with a small offset and large luminosity gap 
indicates a dynamically-relaxed, merger-free system; such systems 
are an ideal testbed for studies of the connection between galaxy
 interactions and AGN activity. Following the results obtained by
Raouf et~al., Khosroshahi et~al.\ (2017) showed that
the relaxed galaxy groups selected by this method are under luminous in radio
compared to the unrelaxed groups. We wish to confirm that 
the results presented in Section \ref{sec:RR} are not biased by these offsets.

We note that the FOF and other optically-based group
finding algorithms can not characterize the dynamical status of groups to 
select virialized and relaxed systems. Therefore, additional constraints are frequently
applied. Pearson et al. (2017) apply a limit on the richness and use 
a set of substructure tests to find relaxed groups. Their method is 80 percent
reliable but dramatically filters out the FOF group sample.
The success of the Raouf et~al.\ method in finding relaxed groups is based on 
hydrodynamical simulations. In observational studies, high reliability in selecting 
virialised and relaxed groups can only be reached with the detection of X-ray emission 
of the hot halo gas, which will be available in future X-ray surveys.

In the absence of group X-ray data, we follow Khosroshahi et~al.\ (2017) in 
defining the centre of the group as the luminosity-weighted centre. 
We use the N$\geq$4 groups for this analysis to ensure enough group members 
to allow a sensible position estimate. Galaxy group samples (with 
both M$\rm_{r}$$\leq$-20 and M$\rm_{r}$$\leq$-19 cuts) are then divided into those cases where the 
BGG lies within 100 kpc of the group centre, and those where the 
offset is larger than 100 kpc. Within each subset, the fraction of 
radio AGN versus luminosity gap has been calculated and is presented in Fig \ref{figoff}. 
We find no correlation between luminosity gap and the fraction 
of sources that are radio-bright, for either small or large offset BGGs.
This result holds for both the M$\rm_{r}$$\leq$-20 sample, where we have higher S/N 
at lower luminosity gaps and for the  M$\rm_{r}$$\leq$-19 sample where we have lower S/N
but can probe higher luminosity gaps.
Note that the small decline in the highest mass bin  ($\Delta$M$\rm_{r}$$\sim$2) for the 
M$\rm_{r}$$\leq$-20 sample is not statistically significant, and is not seen in the N$\geq$2 and 3 samples. 
Here, the small sample sizes at large luminosity gap bins prevent us from examining whether the results hold if a higher radio luminosity limit is also applied, as we did in Section 5. 

The result of this section for large luminosity gap systems ($\Delta$M$\rm_{r}$$>$1.7)
is not consistent with that of Khosroshahi et al. (2017) which reports a lower radio luminosity for BGGs with small offset to the
center of large luminosity gap groups than those with large offset to the center of small luminosity gap groups. 
Their results may be affected by redshift bias as there is a significant difference
between the redshift distributions of the small and large luminosity gap groups in their study, and their study did not take account of the radio survey completeness limit.  Since their study covered a relatively wide range of redshifts (0$<$z$<$0.3),
using incomplete samples may affect the results significantly.

The -20 cut sample shows indications of a higher fraction 
of radio-loud BGG for the groups with small offset than those with large offset. 
The differences are significant in the stellar mass-magnitude gap 
bins which we have higher signal to noise ratio. 
Differences of $\sim$2$\sigma$ have been observed for the low mass BGGs at $\Delta$M$\rm_{r}$$\sim$0.1, 
middle mass BGGs at $\Delta$M$\rm_{r}$$\sim$0.8 and massive BGGs 
at $\Delta$M$\rm_{r}$$\sim$1.4 in the left-hand panel of Fig 7.
This result is consistent with 
Best et~al.\ (2007), who find a higher fraction of radio AGN for the galaxies within
0.2$r\rm_{200}$ in galaxy clusters. 
The difference is not seen for the -19 cut sample in which a lower 
radio luminosity cut has been used. This suggests that the very bright radio AGN can only be found 
in the group centre, where the accretion rates from hot gas halo are likely to be higher.

\section{Summary and conclusions}
\label{sec:summery}

\noindent
We have investigated the radio and optical properties of a sample of 
galaxy groups from SDSS in order to study the effect of
galaxy interactions on the AGN activity within galaxy groups using
 the group luminosity (or magnitude) gap. No variation has been detected
for the fraction of radio loud BGGs or for any of the optical properties of the BGGs 
with respect to the magnitude gap up to 
$\Delta$M$\rm_{r}$=2.7. 
We study the effect of BGG offset from the center of group on the 
AGN activity-magnitude gap results for 
the BGGs in a 100 kpc radius from the center and outside this radius
and show that the radio-AGN fraction is higher for the 
BGGs located close to the group centre. However no sign of correlation
with luminosity gap has been detected for either sample.

Construction of a large observational sample of systems with
large $\Delta$M$\rm_{r}$ is not possible based on 
current optical and X-ray surveys. Instead, we investigate large $\Delta$M$\rm_{r}$ systems  
using the state-of-the-art EAGLE hydrodynamical simulations. The simulation data
make it possible to explore the dependence of BGG properties on the magnitude gap 
up to $\Delta$M$\rm_{r}$$\sim$4. We replicate the observational 
results and find no change in the optical properties of the BGGs
up to the highest $\Delta$M$\rm_{r}$. This suggests that we aren't constrained by 
the observational limits, although the radio result 
still needs more investigation at large $\Delta$M$\rm_{r}$.

We conclude that the AGN activity, as traced by the radio emission,
does not change as the magnitude gap in the group evolves up to 
$\Delta$M$\rm_{r}$=2.7. The black hole power 
is more influenced by the BGG properties such as 
stellar mass than the group properties when the magnitude gap
is considered. In a recent study, Sabater et~al.\ 2019 also found that they were more 
dependent on stellar mass than on black hole mass and that 
therefore stellar mass appears to be the driving factor. 
Any conclusion above $\Delta$M$\rm_{r}$=2.7 limit 
will need a significant increase in the sample size
or deeper optical and radio observations.

\section*{Acknowledgments}
\noindent
This work has been supported financially by Research 
Institute for Astronomy \& Astrophysics of Maragha (RIAAM)
and Institute for Astronomy Royal Observatory Edinburgh
via the visitor grant. HM would like to thank Dr Farhad Daei
for his generous technical support. PNB and JS are grateful for 
support from the UK STFC via grant ST/R000972/1. RKC acknowledges funding from 
an STFC studentship and the John Harvard Distinguished Science Fellowship. 
We acknowledge the Virgo Consortium for making
their simulation data available. The EAGLE simulations
 were performed using the DiRAC-2 facility at
Durham, managed by the ICC, and the PRACE facility
 Curie based in France at TGCC, CEA, Bruy{\'e}resle-Ch{\^a}tel.

\appendix
\section{The EAGLE Data SQL query}

SELECT\\
\indent             FoF.GroupID as groupid,\\
\indent             FoF.GroupMass as halomass,\\
\indent             AP\_Star.Mass\_Star as sm,\\
\indent             SH.BlackHoleMass as bhm,\\
\indent             SH.SubGroupNumber as groupno,\\
\indent             SH.SnapNum as snapshotno,\\
\indent             SH.InitialMassWeightedStellarAge as stellarage,\\
\indent             SH.HalfMassRad\_Star as r50,\\
\indent             SH.HalfMassProjRad\_Star as r50\_proj,\\
\indent             AP\_Star.SFR as sfr,\\
\indent             AP\_Star.Mass\_Gas as gasmass,\\
\indent             AP\_Star.Mass\_BH as apbhm,\\
\indent             magtable.g\_nodust as gmag\_nodust,\\
\indent             magtable.r\_nodust as rmag\_nodust,\\
\indent             dustmagtable.SDSS\_r as rmag\_dust\_sdss,\\
\indent             dustmagtable.SDSS\_g as gmag\_dust\_sdss,\\
\indent             dustmagtable.SDSS\_r\_e as rmag\_dust\_sdss\_edgeon,\\
\indent             dustmagtable.SDSS\_g\_e as gmag\_dust\_sdss\_edgeon,\\
\indent             dustmagtable.SDSS\_r\_f as rmag\_dust\_sdss\_faceon,\\
\indent             dustmagtable.SDSS\_g\_f as gmag\_dust\_sdss\_faceon,\\
\indent             SH.MassType\_Star as SH\_M\_stellar,\\
\indent             SH.MassType\_BH as SH\_M\_BH,\\
\indent             Size.R\_halfmass30 as r50\_30,\\
\indent             Size.R\_halfmass30\_projected as r50\_30\_pro\\
FROM\\
\indent             RefL0100N1504\_Subhalo as SH,\\
\indent             RefL0100N1504\_Aperture as AP\_Star,\\
\indent             RefL0100N1504\_FOF as FoF,\\
\indent             RefL0100N1504\_Magnitudes as magtable,\\
\indent             RefL0100N1504\_DustyMagnitudes as dustmagtable,\\
 \indent            RefL0100N1504\_Sizes as Size\\
WHERE\\
\indent             SH.SnapNum = 28\\
\indent             and SH.GalaxyID = AP\_Star.GalaxyID\\
\indent             and SH.GalaxyID = magtable.GalaxyID\\
\indent             and SH.GalaxyID = dustmagtable.GalaxyID\\
\indent             and SH.GroupID = FoF.GroupID\\
\indent             and SH.GalaxyID = Size.GalaxyID\\
\indent             and AP\_Star.ApertureSize = 30\\
\indent             and AP\_Star.Mass\_Star $>$ 1e9\\

\end{document}